%% file: Survey_LLM_Prompting.tex
\documentclass[journal]{IEEEtai}

\usepackage{xcolor,soul,framed} 

\colorlet{shadecolor}{yellow}

\usepackage[pdftex]{graphicx}
\graphicspath{{../pdf/}{../jpeg/}}
\DeclareGraphicsExtensions{.pdf,.jpeg,.png}

\usepackage[colorlinks=true,
            linkcolor=blue,
            citecolor=blue,
            urlcolor=cyan]{hyperref}
\usepackage[cmex10]{amsmath}

\usepackage{array}
\usepackage{mdwmath}
\usepackage{mdwtab}
\usepackage{eqparbox}
\usepackage{makecell} 
\usepackage{url}
\usepackage{booktabs}
\usepackage{pifont}
\usepackage{subfigure}
\usepackage{multirow}

\usepackage{tikz,xcolor,xparse,booktabs,array}
\usetikzlibrary{calc}

\definecolor{cw1}{RGB}{238,199,159}     
\definecolor{cw2}{RGB}{241,223,164} 
\definecolor{cw3}{RGB}{116,182,159}   
\definecolor{cw4}{RGB}{166,205,228}   
\definecolor{cw5}{RGB}{226,200,216} 

\definecolor{image1}{RGB}{203,216,240} 
\definecolor{image2}{RGB}{247, 198, 204} 
\definecolor{image3}{RGB}{181, 237, 232} 
\definecolor{image4}{RGB}{255, 240, 166} 

\definecolor{email1}{RGB}{237, 247, 253} 
\definecolor{email2}{RGB}{254, 247, 233} 
\definecolor{email3}{RGB}{217, 237, 209} 

\usepackage{adjustbox}

\newcommand{\scorecolorname}[1]{
  \ifnum#1<3 cw1\else
  \ifnum#1<5 cw2\else
  \ifnum#1<7 cw3\else
  \ifnum#1<9 cw4\else cw5\fi\fi\fi\fi
}

\NewDocumentCommand{\scorecell}{ O{0.18} O{0.06} O{0} m m m m m }{
  \begin{tikzpicture}[baseline=0.6ex, x=#1cm, y=#1cm]
    \pgfmathsetmacro{\step}{1 + (#2/#1)} 
    \foreach \v [count=\i from 0] in {#4,#5,#6,#7,#8}{
      \pgfmathsetmacro{\x}{\i*\step}
      \edef\cname{\scorecolorname{\v}}
      \fill[\cname,rounded corners=0.06] (\x,0) rectangle ++(1,1);
      \draw[rounded corners=0.06]        (\x,0) rectangle ++(1,1);
      \ifnum#3=1
        \node[font=\scriptsize] at (\x+0.5,0.5) {\v};
      \fi
    }
  \end{tikzpicture}
}

\makeatletter
\@IEEEtriggercmd{\global\def\@IEEEmaxnumnamesbreadth{2}} 
\@IEEEtriggercmd{\global\def\@IEEEmaxnumnamesdepth{2}}    
\makeatother

\begin{document}

\bstctlcite{IEEEexample:BSTcontrol}

\author{Yiqun~Zhang,~\IEEEmembership{Senior Member,~IEEE,} Yunfan~Zhang, Mingjie~Zhao, Sen~Feng, and~Yiu-ming~Cheung,~\IEEEmembership{Fellow,~IEEE}
\iffalse
\thanks{Received November 2025, revised March 2026, accepted May 2026. This work was supported in part by the National Natural Science Foundation of China (NSFC) under grants: 62476063, the NSFC/Research Grants Council (RGC) Joint Research Scheme under grant: N\_HKBU214/21, the Natural Science Foundation of Guangdong Province under grants: 2025A1515011293 and 2023A1515012855, the General Research Fund of RGC under grants: 12202025 and 12202924, the Guangdong and Hong Kong Universities ``1+1+1'' Joint Research Collaboration Scheme with grant: 2025A0505000004.}
\fi
\thanks{Yiqun Zhang and Sen Feng are with the School of Computer Science and Technology, Guangdong University of Technology, Guangzhou, China 
(e-mail: yqzhang@gdut.edu.cn, 2112305084@mail2.gdut.edu.cn).}
%(e-mail: yqzhang@gdut.edu.cn, 2112305084@mail2.gdut.edu.cn).}
\thanks{Yunfan Zhang, Mingjie Zhao, and Yiu-ming Cheung are with the Department of Computer Science, Hong Kong Baptist University, Hong Kong SAR, China (e-mail: \{csyfzhang, mjzhao, ymc\}@comp.hkbu.edu.hk).} 
\thanks{\textit{Corresponding author: Yiu-ming Cheung.}}
}

\title{How to Ask the AI: A User Perspective Survey for Large Language Model Prompting}

\markboth{IEEE TRANSACTIONS ON ARTIFICIAL INTELLIGENCE
%, VOL. XX, No. XX, MAY 2026
}{Survey for LLM Prompting}

\maketitle

\begin{abstract}

AI tools like ChatGPT and DeepSeek, powered by Large Language Models (LLMs), allow users to obtain instant and effective content responses simply by typing requests, such as ``plan a three-day Vienna trip'', ``solve the attached mathematical problem'', ``draft an email to inquire review progress'', etc., which are also known as LLM prompts. Crafting clear and well-structured prompts leads to more appropriate LLM feedback, which effectively bridges human-LLM interaction. Although prompting appears accessible to non-expert users, precisely organizing effective prompts is a highly systematic and skillful process, presenting potential challenges even for experienced users. This survey explores the principles, taxonomy, and organization of prompts from a user-centered perspective. Differing from the existing surveys that primarily focus on technical principles and application scenarios of LLMs, this paper provides actionable guidelines for formulating effective LLM prompts across diverse real-world tasks and specifically contributes by: 1) developing an intuitive evaluation strategy for prompt efficacy, 2) providing prompting workflow demonstrations on representative applications, and 3) maintaining a dynamically updated open-source project to ensure the core takeaways remain up-to-date. These measures lower the threshold for users to correctly understand and craft prompts that align with evolving application scenarios. This work will be maintained as a living GitHub project \href{https://github.com/Yunfan-Zhang/TAI_Guideline-Table}{\textcolor{blue}{here}}. 
\end{abstract}

\begin{IEEEImpStatement}
This survey bridges a critical gap between the academic and application of LLMs by shifting the focus from technical prompting tuning research to user-centered natural language prompting guidelines. The paper facilitates human-AI interaction for non-specialists while simultaneously contributing application and user-perspective insights to inspire the research domain. By categorizing prompting strategies into organizational (e.g., Q\&A, translation) and creative (e.g., reasoning, content generation) tasks, actionable prompting templates are provided to reduce trial-and-error for LLM users. This work advances the LLM application community in the following aspects: 1) \textit{Democratizing AI} by lowering the barrier to effective LLM use through intuitive, task-specific prompting frameworks; 2) \textit{Enhancing productivity} by offering reproducible workflows for diverse applications, from coding to creative writing, validated via cross-model mutual evaluations; and 3) \textit{Maintaining open-source repository} to keep pace with evolving LLM capabilities and application transitions. Consequently, this survey empowers professionals, including educators, developers, scientists, and practitioners in the entire AI for Science field, to harness LLMs efficiently. The findings and prospects may also offer valuable implications for future human-AI interaction design and even LLM architecture construction.

\end{IEEEImpStatement}

\begin{IEEEkeywords}
User prompting, Large Language Models, LLMs, taxonomy, survey, AI for science, prompting process.
\end{IEEEkeywords}

\IEEEpeerreviewmaketitle

\input{TPLLM_Tex_File/1_Introduction}

\input{TPLLM_Tex_File/2_Taxonomy}

\input{TPLLM_Tex_File/3_Application}

\input{TPLLM_Tex_File/4_exp}

\section{Conclusion}
\label{sct: conclusion}

\textbf{Summarization:} This survey systematically examined prompting techniques for LLMs from users' perspectives. It categorizes prompting strategies by the organizational and creative task types, evaluates their effectiveness through a novel mutual evaluation mechanism to ensure quantification and fairness, and visually demonstrates the prompting process and feedback through a variety of case studies. Key findings reveal that: 1) direct prompting (e.g., zero-shot, persona-based) excels in simplicity and speed for routine tasks, while reasoning-based ones (e.g., chain-of-thought) enhance complex problem-solving; 2) context and constraints in prompts significantly improve output quality, as seen in the case study of storytelling and image generation applications; and 3) no one-size-fits-all approach exists, i.e., optimal prompting depends on the demands of tasks and the goals of users.

\textbf{Prospect of prompting:} Despite the contribution of this work, it is also worth discussing the future development trend and risks of prompting techniques: 1) \textit{Automatic prompts optimization} based on user behavior can reduce trial-and-error. However, excessive automation could diminish user control and, in the long term, potentially impair the development of human cognitive abilities; 2) \textit{Multimodal prompts integration} that combines text, voice, and visual prompts could unlock richer interactions. Nevertheless, biases and misinterpretations dominated by specific modalities will be difficult to eliminate due to the natural gaps between modalities; 3) \textit{Research into prompt engineering} reveals the mechanisms of LLM-prompt interactions, supporting the crafting of customized and functionally diverse prompts to enhance LLM efficacy. However, this also lowers the barrier to ``jailbreaking'' of LLMs, potentially leading to harmful outputs. Consequently, mastery of prompt techniques lowers the threshold for using AI to enhance productivity and increase the flexibility of LLM usage. Yet it simultaneously introduces more uncertainty into already complex and unstable LLMs, elevating ethical concerns and risks of malicious use. Therefore, user education on safety and the development of attack prevention mechanisms for large models become imperative.

\textbf{Promising future research orientations:} Building upon the aforementioned developmental prospects of LLM prompting, several high-potential research directions are emerging: 1) trustworthy adaptive prompting models, 2) multimodal prompt fusion, and 3) prompt-based attack and defense techniques. By systematically addressing these challenges, prompting can become a more reliable interface for human-AI collaboration.

\ifCLASSOPTIONcaptionsoff
  \newpage
\fi

\bibliographystyle{IEEEtran}
\bibliography{Survey_LLM_Prompting}

\end{document}

%% file: TPLLM_Tex_File/1_Introduction.tex
\section{Introduction}\label{sct:intro}
\IEEEPARstart{A}{rtificial} Intelligence (AI) has shown significant advancement with the development of LLMs~\cite{roberts2024artificial, mao2025llms} like ChatGPT~\cite{LMFL, openai2023gpt4} and DeepSeek~\cite{guo2025deepseek}, which are trained on massive corpora of text documents~\cite{LLMSurvey1, minaee2024large}. With continuous improvements, LLMs have been refined to capture deeper linguistic structures and contextual information~\cite{wang2025parameter}, thus enhancing their ability to generalize across tasks and domains, e.g., knowledge Question \& Answering (Q\&A)~\cite{agrawal2024cyberq}, general conversation~\cite{abbasiantaeb2024let}, code generation~\cite{wang2025teaching}, and research assistance~\cite{liu2025towards} for general users, scientists, and AI researchers. Given advanced LLMs, e.g., ChatGPT and Gemini, the most common way of interaction for users is through natural language instructions, also known as prompting\footnote{In this survey, the term prompting mainly refers to natural-language instructions given by end-users, while `prompt tuning' that involves functional adjustments of LLMs will also be briefly introduced in Section~\ref{sct: prompt_tax}.}~\cite{vatsal2024survey, fagbohun2024empirical}, which provides numerous freedoms for users to utilize LLMs~\cite{Zhang2025AI}.

\begin{figure}[!t]
    \centering
    \includegraphics[width=1.02\linewidth]{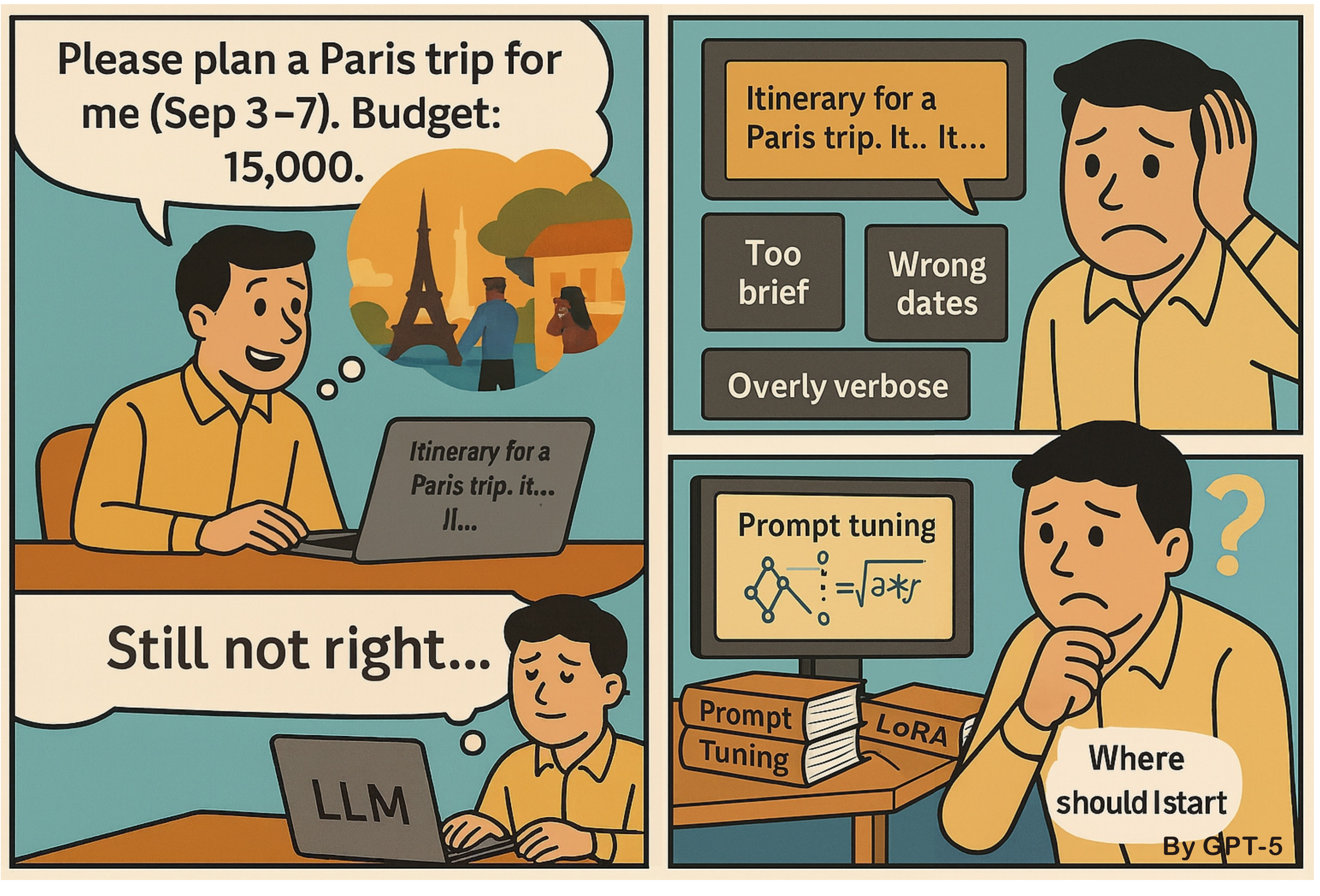}
    \caption{Ineffective human-LLM interaction caused by limited prompting skills and learning barriers.}
    \label{fig:prompting-importance}
\end{figure}

Although prompting appears accessible to non-expert users, precisely organizing effective prompts is a highly systematic and skillful process, presenting potential challenges even for experienced users. Studies on user experiences of prompting~\cite{braun2024can} suggest that the difficulties often lie in three typical dilemma scenarios: 1) Felt that LLMs are powerful but hard to control, always struggling to get LLMs to produce exactly what they need; 2) Confused by the required length and detail for prompts, as well as what kind of information must be included for LLMs to properly comprehend the intention; 3) Every time prompting with LLMs feels like starting from scratch due to the lack of a ``replicable template'', with trial and error dominating the process~\cite{Zamfirescu2023prompt}, as shown in Fig.~\ref{fig:prompting-importance}. Therefore, structural, repeatable, and application scenario-linked prompt organization templates are urgently needed. Accordingly, within the systematic investigation of LLM prompting, the following four issues are imperative: 1) Why is prompt design needed? 2) Which type of prompting is suitable for certain tasks? 3) How to organize good prompts? 4) What does the prompting process look like?

\iffalse
\begin{itemize}
    \item[\textbf{Q1:}] Why prompting? (Sections~\ref{sct:intro})
    \item[\textbf{Q2:}] Which prompt is suitable? (Section~\ref{sct:taxonomy})
    \item[\textbf{Q3:}] How to write good prompts? (Section~\ref{sct: application})
    \item[\textbf{Q4:}] What does prompting process look like? (Section~\ref{sct: exp}) 
\end{itemize}
\fi

To elucidate the necessity of prompt design, it is essential to first understand the fundamental interaction mechanism between LLMs and human users. As LLMs continue to advance, they have transcended their initial roles as simple conversational agents, developing into sophisticated systems capable of semantic comprehension and content generation, while being able to exhibit capabilities akin to human reasoning when solving complex tasks~\cite{muktadir2023brief}. Accordingly, meticulously crafted prompts are required to implicitly map user intentions to the actual inference processes of LLMs~\cite{marvin2023prompt, kraljic2024prompt}. That is, user prompts undergo segmentation~\cite{chen2025unleashing}, specific encoding~\cite{shang2025pro}, and structural analysis, ultimately activating the model’s response. However, given that LLMs remain complex closed boxes to users, the interaction between humans and LLMs fundamentally differs from human-to-human communication, often resulting in unexpected or suboptimal outputs. This gap necessitates treating prompts not merely as input strings, but as an intermediary language for interacting with LLMs~\cite{sahoo2024systematic, chen2023prompting}. Thus, prompt engineering has emerged as a critical skill to bridge the gap between human cognitive patterns and the operational logic of LLMs to unlock their full potential while mitigating the unpredictability of their outputs.

In daily prompting practice, it can be commonly observed that the same prompt can yield different responses from an LLM across multiple attempts, and even slight rephrasing of the prompt may lead to varied outputs~\cite{errica2024did}. From the LLM's perspective, when a user provides a prompt, the LLM first interprets the text into a form it can process internally. Based on its training on vast amounts of text, the LLM then predicts what words are most likely to follow, until it forms a complete sentence-wise or paragraph-wise response~\cite{gpt2}. In simple terms, LLMs do not attempt to ``understand'' human prompts. Instead, through the patterns of language, reasoning, and information learned from numerous training corpora, they can search their embedded knowledge according to the prompts and organize the corresponding answers in the form of human language~\cite{Kojima2022zerocot, chen2021evaluating}. This complex processing pipeline introduces a certain degree of non-intuitiveness and uncertainty into human-LLM interactions.

Empirical evidence suggests that incorporating more detailed specifications in prompts, applying appropriate constraints, or engaging in multi-round dialogue for refinement can significantly enhance output quality. For example, asking ``What's the best time of year to visit Paris?'' might get a hasty and general answer, as shown in Fig~\ref{fig:prompting-importance}. By contrast, adding persona context like ``Assume you are a travel guide and I love off-seasons. In simple terms, What's the best time to visit Paris and why?'' is more likely to yield a response that is closer to the actual needs of users. From experience, by adopting prompting strategies such as giving context, setting constraints, or breaking a task into steps, with their technical principles explained in the literature~\cite{an2024make},
%~\cite{zhang2024rest}, and~\cite{zhang2024rest}, 
\cite{zhang2024rest}, users can guide LLMs to generate more consistent results~\cite{carta2023grounding}. Nevertheless, mastering these isolated experiential techniques demands prolonged practice, while determining suitable prompting strategies for varying use cases remains a persistent difficulty for users. These challenges underscore the necessity for a survey to demystify prompting and provide systematic prompting manuals. 

In the literature, most surveys focus more on technical characteristics of LLMs, such as theoretical principles~\cite{huang2024survey, yang2024model, chang2024efficient, liu2023pre} and benchmark performance~\cite{mohammadi2025evaluation, li2025benchmark, chang2024survey, ni2025survey, jain2025systematic}. Although a previous survey~\cite{zhang2025trending} introduces LLMs from an ordinary user perspective, it mainly reveals task-LLM matching. Following the user-centered style of~\cite{zhang2025trending}, this survey elaborately discusses the main types of prompting strategies, their applicable cases, and organization. To help users quickly locate appropriate prompting strategies, the use scenarios are divided into organizational and innovative tasks. The organizational scenarios emphasize information management, e.g., conversation, or knowledge Q\&A, where short and direct prompts may suffice to get a good answer~\cite{wang2024prompt, amatriain2024prompt}. For more complex innovative tasks for solving complex math problems, drafting structured content, etc., users often benefit from adding details to prompts or breaking them into multiple steps~\cite{wei2022chain,baldassini2024makes}. To provide readers with an intuitive understanding of prompting capabilities, this paper employs the LLM judges~\cite{li2024llms, wei2024systematic} to quantify the capabilities of various prompting strategies across different application scenarios. Furthermore, demonstrations of prompting workflows in benchmark tasks are included, closing the last mile in making scientific prompting accessible to general users. The main contributions of this survey can be summarized into four-fold:
\begin{itemize}
\item \textbf{Users centered narrative.} This survey represents the first review of prompting techniques from a non-AI researcher's perspective. By systematically examining the role of prompts across diverse routine tasks, it lowers the entry barrier, enabling readers to quickly grasp prompting and leverage the power of LLMs.

\item \textbf{Task-oriented taxonomy.} Prompting strategies are categorized by task types: 1) organizational tasks: conversation, Q\&A, language translation, and 2) innovative tasks: content generation, reasoning, and coding. This helps readers easily map their specific LLM usage needs to appropriate prompting strategies.

\item \textbf{Intuitive demonstrations.} An intuitive quantitative evaluation system and a set of practical task demonstrations are designed to showcase how different prompting styles perform in specific scenarios. The original prompts in the tasks are also demonstrated, closing the final replication gap in real-world applications.

\item \textbf{Sustainability.} The core content of this survey is designed to be continuously updated: Taxonomy, prompting templates, key examples, and case studies will be maintained as an open GitHub project, ensuring that the audience of this paper keeps up with timely and reliable prompting knowledge and guidance.
\end{itemize}

The organization of the remainder of this paper is as follows: Section~\ref{sct:taxonomy} presents a taxonomy of prompting, offering a cheat sheet for prompting-scenario matching; Section~\ref{sct: application} provides a user prompting manual, showing numerous ``replicable template'' for users in various tasks; Section~\ref{sct: exp} quantitatively evaluates prompting strategies, demonstrates common application cases, showcasing the diverse influence of different prompts; Finally, we concludes the paper and prospects of LLM prompting in Section~\ref{sct: conclusion}.

%% file: TPLLM_Tex_File/2_Taxonomy.tex
\begin{table*}[!t]
\centering
\caption{Taxonomy of LLM Prompting from User Perspective.}
\label{tab:prompting-taxonomy}
\renewcommand{\arraystretch}{1.3} 
\setlength{\tabcolsep}{3pt}       
\resizebox{2.08\columnwidth}{!}{
\begin{tabular}{>{\centering\arraybackslash}m{3cm}|
                >{\centering\arraybackslash}m{5cm}|
                >{\centering\arraybackslash}m{7cm}|
                >{\centering\arraybackslash}m{5cm}}
\toprule
\textbf{Category} & \textbf{Representative Approaches} & \textbf{Principle} & \textbf{Limitations} \\
\midrule

Direct Prompting & Zero-/Few-shot~\cite{LMFL}, Persona~\cite{schick2021s}, Instruction~\cite{mishra2021reframing} Prompting  &
Provide task description and a few examples or explicit instructions to guide model output. &
Sensitive to template wording and example order. \\
\midrule

Reasoning-based & Prompt chaining~\cite{wei2022chain}, Chain-of-thought~\cite{wei2022chain, guo2025deepseek}, Tree-of-thought~\cite{yao2023tree}, Least-to-most~\cite{zhou2022large}&
Encourage step-by-step reasoning or decompose problems into sub-tasks. &
Higher inference cost; requires large models. \\
\midrule

Retrieval/Tool-Augmented & Retrieval-Augmented Generation (RAG)~\cite{lewis2020retrieval}, Reasoning and Acting (ReAct)~\cite{yao2023react} &
Combine prompts with retrieval or external tools to reduce hallucination and support factual accuracy. &
Dependence on external systems. \\
\midrule

Ensemble-based & Context Calibration~\cite{zhao2021calibrate}, Multi-template Voting~\cite{gou2023mvp}&
Generate multiple response and aggregate them for stability. &
Additional calibration cost. \\
\midrule

Self-improve & Self-consistency~\cite{wang2023selfconsistency}, Self-refine~\cite{madaan2023self} &
Iterative refine outputs to enhance performance. &
Increased computation overhead. \\
\midrule

Agent-based & AutoGPT~\cite{auto_gpt}, BabyAGI~\cite{babyagi}, LangChain Agents~\cite{langchain}  &
Orchestrate multi-step actions by combining planning, tool-use, and memory, driven by prompts. &
Complex to design; unstable or costly in real-world use; risk of privacy leakage.\\
\midrule

Learnable Prompts & Prompt Tuning~\cite{ding2023parameter,liu2022p}, Prefix Tuning~\cite{li2021prefix} &
Freeze model and train lightweight prompt vectors for task adaptation. &
Require task-specific training data. \\
\midrule

Automated Construction & AutoPrompt~\cite{shin2020autoprompt}, Generated Knowledge Prompting~\cite{zhou2022large}&
Automatically search or generate prompt templates and examples. &
Higher implementation complexity. \\
\bottomrule
\end{tabular}}
\end{table*}

\section{Prompting Taxonomy}\label{sct:taxonomy}

Two key questions are addressed in this section: 1) What are the different prompting strategies? 2) What tasks can prompting enable LLMs to do? To answer the first question, the taxonomy of prompting strategies is explored, and different prompting approaches are categorized based on their underlying principles. To address the second question, prompting strategies are discussed under application scenarios of organizational and innovative tasks.

\subsection{Prompting Strategies}
\label{sct: prompt_tax}
Prompting strategies can be broadly categorized into several types according to their mechanisms and principles, which are summarized in Table~\ref{tab:prompting-taxonomy}. Limitations of different prompting strategies~\cite{zhao2023survey,zhou2022large} are also provided to make the table a practical cheat sheet.

As summarized in Table~\ref{tab:prompting-taxonomy}, most categories, i.e., direct, reasoning-based, retrieval/tool-augmented, ensemble-based, and self-improve prompting, can be directly realized through natural-language interaction. The \textit{direct prompting}, including sub-types of zero-shot, few-shot~\cite{LMFL}, instruction~\cite{mishra2021reframing} and persona prompting~\cite{schick2021s,kong2024better}, guides model outputs through explicit commands or a small number of examples. These strategies are simple and efficient, but they are sensitive to wording and the order of examples. The \textit{reasoning-based prompting}, such as prompt chaining~\cite{wei2022chain}, chain-of-thought~\cite{wei2022chain, guo2025deepseek,wang2023plan}, tree-of-thought~\cite{yao2023tree}, and least-to-most~\cite{zhou2022large} prompting, enhances logical and multi-step problem solving by encouraging intermediate steps. However, these approaches typically increase inference length, thus may require sufficiently large models to provide functional support. The \textit{retrieval/tool-augmented prompting}, exemplified by Retrieval-Augmented Generation (RAG)~\cite{lewis2020retrieval} and Reasoning and Acting (ReAct)~\cite{yao2023react}, enhances response reliability by telling the model to retrieve relevant evidence or tool outputs prior to answering, thus grounding generation on factual context. The \textit{ensemble-based} strategies, such as context calibration~\cite{zhao2021calibrate} and multi-template voting~\cite{gou2023mvp}, focus on aggregating responses to stabilize results but often come with added computational costs or complexity. \textit{self-improve prompting} (e.g., self-consistency~\cite{wang2023selfconsistency}), focuses on generating multiple initial outputs first, then iteratively refining them to enhance performance.

\begin{figure}[!t]
    \centering
    \includegraphics[width=1.02\linewidth]{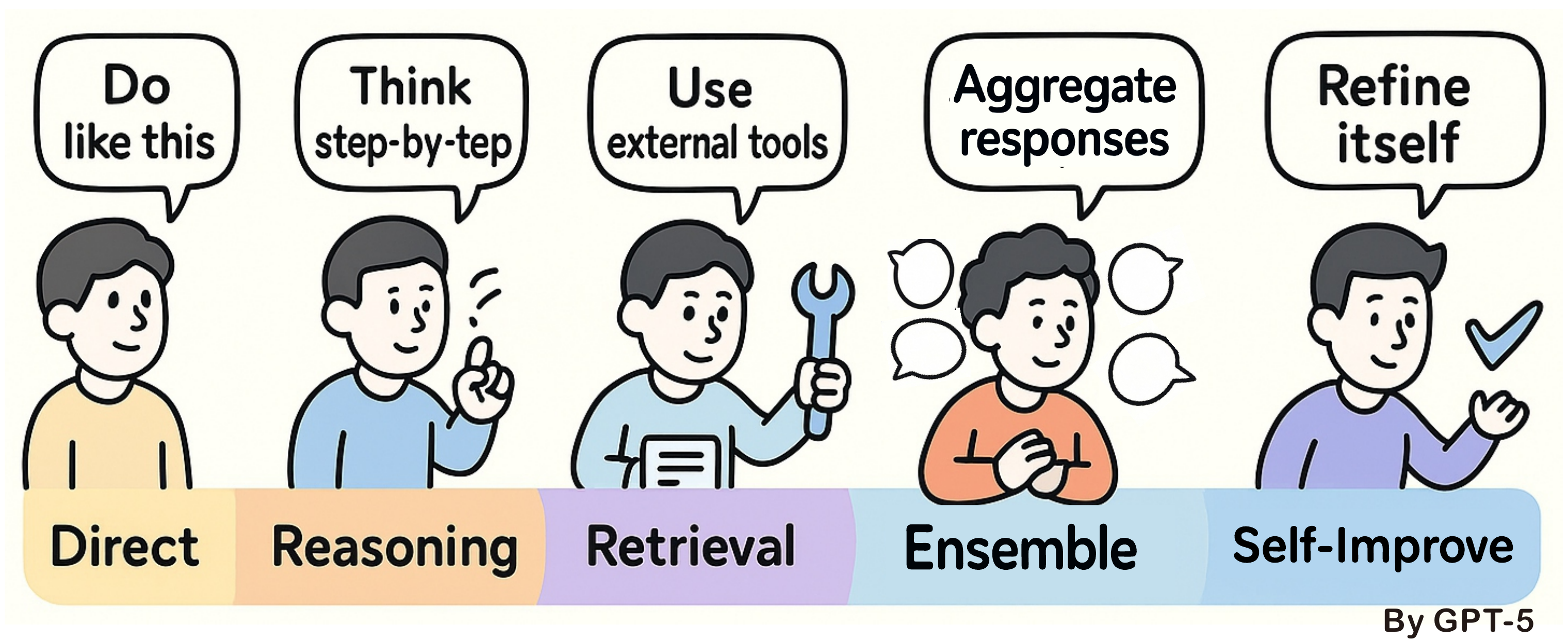}
    \caption{Intuitive comparison of the five natural language prompting strategies: Direct Prompting (Direct), Reasoning-based (Reasoning), Retrieval/Tool-Augmented (Retrieval), Ensemble-based (Ensemble), and Self-improve.}
    \label{fig:prompting-map}
\end{figure}

\begin{figure}[!t]
    \centering
    \includegraphics[width=1.02\linewidth]{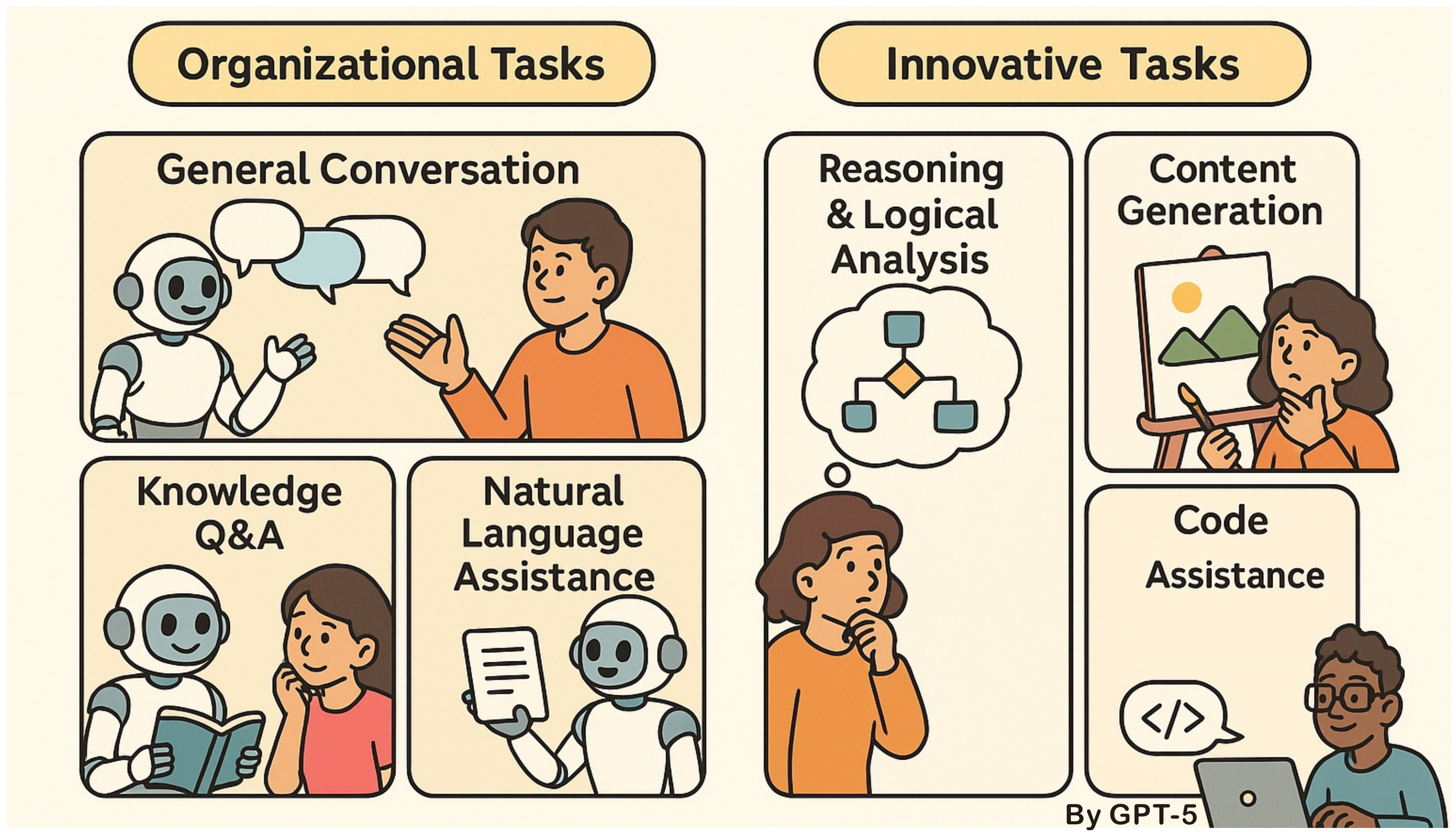}
    \caption{LLM prompting of organizational and innovative tasks.}
    \label{fig:task-map}
\end{figure}

The other prompting categories, i.e., agent-based, learnable prompts, and automated construction, involve mechanisms designed beyond natural language-level prompts. The \textit{agent-based prompting}, such as AutoGPT or LangChain agents~\cite{auto_gpt,langchain}, orchestrates multi-step actions by integrating planning, tool use, and external memory that stores intermediate results and contextual information across reasoning steps. \textit{learnable prompting} (e.g., prompt tuning~\cite{ding2023parameter,liu2022p}, its derived visual prompt tuning~\cite{li2024improving,liu2025fate}, and prefix tuning~\cite{li2021prefix}) optimizes trainable prompt vectors to steer the model toward specific downstream tasks. The \textit{automated prompting} (e.g., AutoPrompt~\cite{shin2020autoprompt}, Generated Knowledge Prompting~\cite{zhou2022large}) employs optimization-based or gradient-guided search to automatically construct effective prompts that elicit desired behaviors from models. A common limitation of these categories is that they require additional training and optimization that are out of reach for ordinary users.

In summary, prompting strategies can be viewed as a continuum: from direct and reasoning-based prompts that rely solely on natural-language interaction, to retrieval-augmented, ensemble-based, and self-improve, that enhance stability and factuality, and further to agent-based, learnable, and automated approaches that extend beyond simple prompt design, as illustrated in Fig.~\ref{fig:prompting-map}. These categories complement each other rather than being mutually exclusive, together forming the foundation of how prompts shape model behavior. Building upon this mechanism-oriented taxonomy, the next subsection turns to an application scenario perspective, outlining how different types of prompts align with and act in fulfilling specific application tasks.

\subsection{Task-Specific Prompting Approaches}\label{sct:tasks}

To answer ``what tasks can prompting enable LLMs to perform?'', the applicability of different prompting approaches introduced in Section~\ref{sct: prompt_tax} is introduced in application scenarios demonstrated in Fig.~\ref{fig:task-map}. From an application perspective~\cite{zhang2025trending}, LLM tasks can be broadly divided into: 1) Organizational tasks, which emphasize information management and productivity, and 2) Innovative tasks, which emphasize reasoning, innovative, and open-ended problem solving~\cite{ding2023parameter,bommasani2021opportunities}. Such a taxonomy is derived from established frameworks in the fields of cognitive science and task analysis~\cite{hollender2010integrating}. In the context of cognitive task analysis, human activities are frequently delineated into routine procedural processes (organizational) and open-ended creative processes (innovative). The organizational tasks inherently align with routine knowledge work~\cite{fonseca2019human}, emphasizing direct information retrieval, logical organization, and procedural execution with clear criteria for success (e.g., summarizing text or fetching factual answers). Conversely, the innovative tasks correspond to creative knowledge generation, which demands divergent thinking, complex logical synthesis, and open-ended exploration without a single predefined solution~\cite{sun2022students} (e.g., writing a novel or solving research problems). The alignment between tasks and prompting approaches is provided in Table~\ref{tab:decision_table_prompt} as a cheat sheet. More detailed justifications and prompting manuals are further presented in Section~\ref{sct: application}.

Organizational tasks can be grouped into three categories: general conversation~\cite{openai2023gpt4}, knowledge Q\&A~\cite{lewis2020retrieval}, and natural language assistance~\cite{LMFL}. 
General conversation focuses on generating coherent, context-aware, and human-like dialogue, where direct prompting strategies such as zero-shot, few-shot, or persona prompting are typically sufficient to maintain conversational consistency and stylistic control.
Knowledge Q\&A aims to provide accurate and domain-specific answers by integrating the model’s internal knowledge with retrieved evidence. Retrieval-Augmented prompting (e.g., RAG or ReAct) is therefore preferred, as it grounds responses in up-to-date sources and reduces hallucination. In addition, ensemble-based strategies like context calibration further improve robustness and consistency when prompts or retrieved contexts vary.
Natural language assistance tasks, including translation, summarization, and text correction~\cite{raffel2020exploring}, rely on more structured prompting schemes such as Chain-of-thought or Self-refine, which enable iterative refinement and improve linguistic precision and coherence.

Innovative tasks can also be grouped into three categories: reasoning and logical analysis~\cite{wei2022chain,wang2023selfconsistency}, content generation~\cite{LMFL,madaan2023self,yang2026bridging}, and code assistance~\cite{chen2021evaluating}. Reasoning addresses complex problems through multi-step inference, where reasoning-based prompting (e.g., Chain-of-thought, Tree-of-Thought, or Least-to-most) effectively decomposes problem structures and enhances interpretability. 
Content generation produces coherent and innovative outputs such as stories or designs. This type of task can be well supported by self-improve prompting, which enables iterative drafting and self-revision to enhance creativity, fluency, and overall text quality.
Code generation automates workflows such as synthesis, debugging, and refactoring, accelerating development across programming languages, which can benefit from reasoning-based and self-refine prompting.

\subsection{Task Sensitivity and Data Privacy Issues}\label{sec:sensity}

It is crucial to recognize that prompt efficacy is inherently sensitive to the specific domain and the nature of the input and output. The relative importance of evaluation metrics depends on the user's target tasks, user needs, and available resources. For example, concerning task characteristics, when a user seeks information about critical domains such as medical advice or financial planning, factual accuracy and reliability mean more to prevent harmful consequences. Conversely, for everyday organizational tasks, such as finding the best local restaurant or summarizing a casual email, efficiency often takes precedence. Regarding user needs, if the target user is an elementary school student requiring straightforward answers, easy-to-use carries a significantly higher weight, ensuring the output is highly accessible without unnecessary prior knowledge and complexity to understand. Finally, in terms of available resources, in scenarios where the provided input information is extremely limited, the interaction should rely largely on the LLM's free generation, where the ability and compliance of LLMs to handle various tasks become the critical indicators.

Furthermore, the application of advanced prompting techniques carries non-negligible data and privacy implications. Strategies like RAG or tool-augmented prompting in Table~\ref{tab:prompting-taxonomy} mitigate hallucinations by anchoring the model to external knowledge bases. However, utilizing these techniques via cloud-based LLM APIs to process sensitive data introduces severe data leakage vulnerabilities. Such risk is particularly catastrophic for users and industries strictly bound by confidentiality, such as those handling proprietary corporate strategies, unreleased financial data, or sensitive medical diagnostics, where any exposure could lead to severe legal and ethical compromises. Therefore, users should carefully weigh the trade-off between the enhanced task performance of augmented prompting and its inherent privacy risks. When handling sensitive information, deploying local models or enforcing strict data anonymization would be promising.

%% file: TPLLM_Tex_File/3_Application.tex
\begin{table}[!t]
\color{black}
\centering
\caption{A decision table guiding users to determine suitable prompting strategies for the corresponding tasks.}
\resizebox{1\columnwidth}{!}{
\begin{tabular}{c|c|c}
\toprule
\multicolumn{2}{c}{\textbf{Tasks}} &  \textbf{Recommended Strategies} \\
\midrule

\multirow{3}{*}{\rotatebox[origin=r]{90}{\makebox[2cm][c]{\ \ \ \ Organizational Tasks}}}&\shortstack{General \\ Conversational} &  \shortstack{Zero-shot,  Few-shot, Persona, \\ Prompt chaining, Multi-template Vote} \\
\cmidrule{ 2 - 3 }
& \makecell[c]{Knowledge \\  Q\&A} & \makecell[c]{Zero-shot,  Few-shot, \\  Retrieval-Augmented Generation, \\  Reasoning and Acting, Context Calibration}  \\

\cmidrule{ 2 - 3 }
&\shortstack{Natural \\ Language Assistance} & \shortstack{Zero-shot,  Few-shot, \\ Chain-of-thought, Self-refine}\\
\midrule

\multirow{3}{*}{\rotatebox[origin=r]{90}{\makebox[1.4cm][c]{\ Innovative Tasks}}}&\shortstack{Reasoning and \\ Logical Analysis} & \shortstack{Chain-of-thought, Tree-of-thought,\\  Least-to-most, Self-consistency} \\

\cmidrule{ 2 - 3 }

&\shortstack{Content \\ Generation} & \shortstack{Zero-shot,  Few-shot, \\ Instruction, Least-to-most}\\

\cmidrule{ 2 - 3 }

&\shortstack{Code \\  Assistance} &  \shortstack{Zero-shot,  Few-shot, \\ Chain-of-thought, Self-refine}\\
\bottomrule
\end{tabular}}
\label{tab:decision_table_prompt}
\end{table}

\section{Prompting Manuals}
\label{sct: application}

\begin{figure}[!t]
    \centering
    \includegraphics[width=1\linewidth]{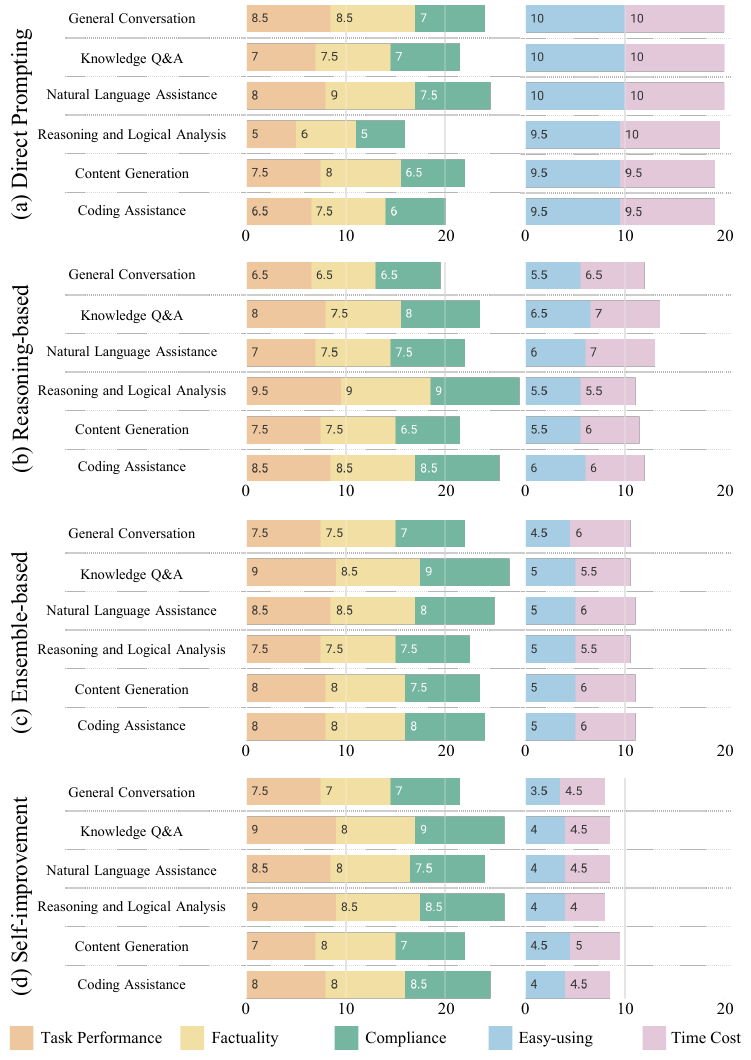}
    \caption{Prompting strategies rated by LLMs across five user-oriented indicators, i.e., \textbf{three effectiveness indicators:} task performance, factuality, and compliance, and \textbf{two efficiency indicators:} easy-using and time cost.}
    \label{fig:table_bar}
\end{figure}

Recent studies report that about 44\% of the experienced LLM users still prefer simple and direct prompts, whereas only around 20\% employ advanced contextualized prompting strategies~\cite{Jin2025understanding}. This tendency toward simple and direct prompting is likely even stronger among ordinary users~\cite{sevigny2025exploring}. Despite their convenience, simple prompting habits fall short when users confront complex, reasoning-intensive scenarios~\cite{shuster2021retrieval}.
Accordingly, this section introduces a detailed usage template of task-specific prompting approaches\footnote{Some of the prompting strategies mentioned in Section~\ref{sct: prompt_tax} are not covered here because they require specialized implementation and cannot be directly used by ordinary users through natural language prompts via the LLM interface.}, including direct, reasoning-based, ensemble-based, and self-improve prompting, under the two task scenarios, i.e., organizational and innovative, introduced in Section~\ref{sct:tasks}. An LLMs-as-judges evaluation is also conducted by asking the LLMs to rate the prompting strategies across different tasks, as shown in Fig.~\ref{fig:table_bar}. Detailed settings and discussion can be found in Section~\ref{sct:cross}.

\subsection{Prompting Strategies for Organizational Tasks}
Organizational tasks emphasize accurate information delivery, logical coherence, and consistency in expression, aiming for outputs that are clear, reliable, and easy to understand~\cite{chen2025unleashing, weir2007cognitive}. These tasks require LLMs to identify information structures, extract key content, and maintain a stable tone, thereby enabling efficient information organization and communication~\cite{zhu2023large, zhang2024comprehensive}. In this process, prompting strategies play a crucial role by providing explicit instructions, format constraints, and stylistic guidance that help the model generate logical, coherent, and dependable outputs, which are introduced under the following three scenarios.

\subsubsection{General Conversation} 
Conversational interactions are common applications of LLMs, where users expect the model to understand context accurately, provide coherent responses, and maintain an appropriate tone throughout~\cite{ deng2025proactive}. The prompting strategies can help shape conversational style, ensure coherence across conversational turns, and make responses sound natural and aligned with user intent~\cite{ein2024conversational}.

\textbf{Zero-shot Prompting:}
In daily conversations, users prefer the direct and concise interaction style of zero-shot prompting~\cite{Kojima2022zerocot}. With this approach, the model can draw on its existing knowledge and generalization ability to complete a task based solely on clear user instructions. This ``ask directly, answer directly'' pattern is both efficient and natural, making it particularly suitable for scenarios that require quick access to factual information. \textit{Example:} ``What is the weather like in Tokyo today?'' Since zero-shot prompting is simple, intuitive, and familiar to most users, only examples are given without discussion when refer to it in different tasks hereinafter.

\textbf{Few-shot Prompting:} 
Rather than giving a single instruction, few-shot prompting provides an LLM with a few examples that illustrate the intended tone or structure, guiding it to follow the same pattern~\cite{blomker2025reevaluating}. By developing a deeper understanding of the provided examples, the model can flexibly adapt to various conversational styles, whether formal, humorous, or narrative. Instead of producing mechanical replies, LLMs can generate semantically coherent and stylistically diverse expressions. \textit{Example:} ``Here is how I like my jokes: `Why did the scarecrow win an award? Because he was outstanding in his field!' Now, tell me another animal joke in a similar style.''

\textbf{Persona Prompting:} 
When users expect a professional perspective rather than a general response, persona prompting serves as a neat option. By guiding the model to assume a specific role or identity, this approach constrains the tone, domain knowledge, and communication style of its outputs, thereby ensuring context relevance~\cite{occhipinti2025harry}. Such targeted role setting effectively narrows the response scope, steering the model away from broad conversational patterns and enabling it to generate more precise and trustworthy responses. \textit{Example:} ``You are a helpful customer support agent for a bank. How can I reset my online banking password?''

\textbf{Prompt chaining Prompting:} 
In real-world applications, users may wish to build upon the responses of LLMs to obtain deeper insights, which requires prompting strategies that effectively maintain conversational continuity. Chained prompting links related prompts and responses into a coherent sequence so that each new reply remains contextually coherent with the preceding dialogue. This approach not only helps prevent the model from ``forgetting'' previous information but also allows the dialogue to progressively deepen its focus on a consistent topic. \textit{Example:} ``Can you elaborate on the second point you just mentioned?''

\textbf{Multi-template Voting Prompting:}
By generating several variations of the same prompt, this strategy produces multiple candidate responses that can be compared or aggregated. In general conversational tasks, it improves reliability by ensuring consistency in both style and information, since the final output reflects agreement across alternatives. Much like asking a question in different ways during a group discussion, this approach makes the answer more robust and trustworthy. \textit{Example:} ```What’s your favorite book?', `Could you recommend a book you enjoy?', `Tell me about a novel you like.' Select the most consistent answer across responses.''

\subsubsection{Knowledge Q\&A}
Retrieving accurate information and presenting it as concise, factual answers lies at the core of knowledge Q\&A. Unlike casual conversation, knowledge-oriented interactions often require synthesizing content from multiple sources, validating consistency across them, and presenting results in an interpretable form. Such tasks typically involve fact-checking, academic inquiries, research assistance, and troubleshooting, where promptings are designed to enhance reliability, reduce factual errors, and guide the model toward providing well-grounded, verifiable responses. 

\textbf{Zero-shot Prompting: } \textit{Example:} ``What is the capital city of Australia?''

\textbf{Few-shot Prompting:}
When questions are ambiguous or require answers in a specific format, few-shot examples provide necessary guidance. By showing the model how similar queries are answered, whether through structured definitions, domain-specific terminology, or stepwise explanations, few-shot examples help produce responses that are more consistent and better aligned with the intended style. \textit{Example:} ``Here are some geography-related examples: `Where is Buenos Aires?' $\rightarrow$ `South America, Argentina', `Where is Barcelona' $\rightarrow$ `Europe, Spain'. Now, answer `Where is Beijing'.''

\textbf{Retrieval-Augmented Generation (RAG) Prompting: } 
Due to the limitations of LLMs in real-time adaptability and the coverage of specialized information, they often struggle to provide accurate answers when dealing with Q\&A tasks involving recent events or domain-specific knowledge. To enhance the credibility of generated content, an effective strategy known as RAG guides the model through prompting to leverage external knowledge sources. RAG enables the model to actively retrieve relevant text segments before generating an answer and to integrate the retrieved evidence into its final output. Such a retrieval and verification mechanism anchors the response in traceable information, thereby significantly reducing the likelihood of hallucination. \textit{Example:} ``Using this product documentation, explain the warranty policy for international customers.''

\textbf{Reasoning and Acting (ReAct) Prompting:} 
Unlike the RAG approach that passively depends on externally retrieved materials, ReAct prompting guides the model to take initiative in seeking information. Instead of being provided with references in advance, the prompt instructs the model to decide when additional evidence is needed and to perform actions such as searching or calculating before continuing its reasoning. This dynamic process allows for a combination of reasoning and information gathering, producing answers that are both accurate and transparent. \textit{Example:} ``Find the current population of Canada. If the information may be outdated, reason what data is needed, search for updated statistics, and summarize your final answer.''

\textbf{Context Calibration Prompting: } 
Even with advanced retrieval and reasoning abilities, LLMs may produce responses that vary in certainty, level of detail, or stylistic consistency. Context calibration prompting addresses this issue by instructing the model to adjust its output based on explicit confidence cues or structural constraints. This approach improves the stability and interpretability of knowledge Q\&A responses, allowing users to more effectively assess the reliability of a given answer. \textit{Example:} ``Answer the following question and indicate your confidence as high, medium, or low: `What is the average lifespan of a blue whale?'''

\subsubsection{Natural Language Assistance}
Natural language assistance encompasses a range of language understanding and expression tasks aimed at maintaining consistency of meaning, tone, and contextual intent across different forms of expression. In addition to multilingual translation, it also includes summarization and text rewriting, all serving the goal of clear and accessible information delivery. Prompting strategies proven to be effective for natural language assistance are introduced below.

\textbf{Zero-shot Prompting:} \textit{Example:} ``Translate `How do you say thank you in Japanese?''' or ``Please briefly summarize the content of the news for me.''

\textbf{Few-shot Prompting: } 
For the tasks that require higher accuracy and greater expressive depth, simple instructions are often insufficient to ensure consistent and well-controlled outputs. In such cases, few-shot prompting~\cite{LMFL} provides clear reference information that establishes a standard for expression, effectively reducing ambiguity and stylistic deviation in translation and summarization. It helps the model maintain stability in tone, level of detail, and domain-specific phrasing, while also offering an implicit structural framework for rewriting tasks that guides sentence organization and stylistic choice, resulting in coherent and purpose-driven outputs.
\textit{Example:} ``Here are examples of rewriting academic sentences into a concise form: `The experiment was conducted to verify the proposed method.' $\rightarrow$ `The experiment verified the proposed method.’ Now, rewrite the sentence in a similar concise academic style: `It is important to note that the algorithm performs well on large datasets.'''

\textbf{Chain-of-thought Prompting: } 
Chain-of-thought prompting strengthens natural language assistance by guiding the model to reason through meaning and structure before generating a response. In formal or information-sensitive scenarios, users often expect outputs that are more precise and appropriate in tone. By encouraging deliberate reasoning, it enables the model to produce language that is logically sound, contextually fitting, and aligned with the user’s communicative intent. \textit{Example:} ``First, identify the central idea of the passage. Next, determine the supporting details that explain or expand it. Finally, present them together in a concise and logically structured paragraph.''

\textbf{Self-refine Prompting: }
Unlike chain-of-thought prompting, which strengthens reasoning before generation, self-refine prompting focuses on improving the text after it is produced. In natural language assistance, users often expect not only correct but also well-polished and fluent responses. By prompting the model to review and refine its own output, this approach enhances clarity, precision, and stylistic quality, resulting in more natural and suitable language. \textit{Example:} ``Generate a summary of the paragraph. Then, review your summary and refine it to improve clarity and readability.''

\subsection{Prompting Strategies for Innovative Tasks}\label{sct:innovative}

Innovative tasks involve open-ended and creative problem-solving scenarios where multiple valid solutions may exist, emphasizing reasoning, imagination, and cross-domain integration~\cite{anantrasirichai2022artificial,boden1998creativity}. In such contexts, prompting serves as a key mechanism for guiding LLMs to express creativity and structured thinking. Through carefully designed instructions, prompting can stimulate divergent exploration, organize logical reasoning, and encourage the synthesis of ideas beyond memorized knowledge~\cite{colton2012computational,LMFL}. To finely explore how different prompting strategies perform in innovative scenarios, the innovative tasks are categorized into three subtypes: Reasoning and Logical Analysis, Content Generation, and Code Assistance. Prompting strategies that are commonly applied for each of the above scenarios are introduced below.

\subsubsection{Reasoning and Logical Analysis}

Reasoning and logical analysis tasks require LLMs to solve problems that involve deduction, inference, and multi-step thinking~\cite{kiciman2023causal}. Unlike simple factual retrieval, these tasks often demand the model to articulate intermediate reasoning, evaluate multiple possibilities, and maintain logical consistency throughout the process. Typical scenarios include solving mathematical word problems, addressing logical puzzles, conducting causal analysis, and supporting strategic planning. In such cases, the effectiveness of prompting lies not only in obtaining the final answer but also in shaping the reasoning path, ensuring interpretability and robustness of the results~\cite{xia2024beyond}. 

\textbf{Chain-of-thought Prompting}: Instead of providing an immediate answer, this approach~\cite{wei2022chain} encourages the model to articulate its reasoning process step by step before reaching a conclusion. In reasoning tasks, this strategy is particularly effective, as it reduces hidden errors, improves logical consistency, and allows users to trace the inference path.
\textit{Example:} ``If John is taller than Mary, and Mary is taller than Peter, who is the tallest? Explain your reasoning step-by-step.''

\textbf{Tree-of-thought Prompting:} Tree-of-thought prompting~\cite{yao2023tree} extends chain-of-thought~\cite{wei2022chain,lyu2023faithful,zhang2024chain} by exploring multiple reasoning branches, evaluating their validity, and then selecting the most promising one. This approach is highly beneficial for tasks with inherent ambiguity or multiple potential solutions, such as puzzles or strategic decision-making.
\textit{Example:} ``Given the following scenario [complex problem description], explore three different potential solutions and outline the logical steps for each, then choose the most effective one and explain why.''

\textbf{Least-to-most Prompting}: Rather than reasoning in a single chain, this strategy~\cite{zhou2022large} decomposes a complex problem into a sequence of simpler and dependent sub-problems to solve progressively. The decomposition helps minimize cognitive load and avoid cascading reasoning errors, where mistakes in earlier steps spread through subsequent reasoning, making it especially effective for challenging multi-step reasoning tasks. \textit{Example:} ``A bakery has 150 loaves, sells 40 in the morning, and bakes 60 in the afternoon. How many are left? Break it down and explain step-by-step.''

\textbf{Self-consistency Prompting:} 
Unlike tree-of-thought prompting, which guides path exploration, self-consistency prompting~\cite{wang2023selfconsistency} generates several independent reasoning paths and compares their conclusions to select the most consistent result. By aggregating multiple reasoning trajectories, this approach enhances output reliability and mitigates single-path biases, improving both confidence and reasoning robustness. \textit{Example:} ``Solve this riddle in three different ways and compare your answers to ensure consistency: `I speak without a mouth and hear without ears. I have no body, but I come alive with wind. What am I?' ''

\subsubsection{Content Generation}

Content generation tasks involve producing original and coherent outputs such as stories, poems, articles, scripts, and marketing copy~\cite{cao2023comprehensive}. Unlike reasoning-oriented tasks that focus on logical correctness, content generation emphasizes novelty, stylistic diversity, and alignment with user intent. Representative scenarios include innovative writing, academic drafting, advertising design, and cross-lingual content creation. In such cases, LLMs are required not only to deliver grammatically correct text but also to adapt tone, style, and narrative flow according to task requirements.

\textbf{Zero-shot Prompting:} \textit{Example:} ``Write a short poem about autumn leaves.''

\textbf{Few-shot Prompting:} By supplying a handful of exemplars~\cite{LMFL}, the user constrains style and format so that the model imitates specific literary or rhetorical patterns. It is especially helpful for maintaining a consistent tone in poetry, essays, and marketing copy.
\textit{Example:} ``Here are two haikus about nature: [Haiku 1], [Haiku 2]. Now write a haiku about city life.''

\textbf{Instruction Prompting:} This approach~\cite{mishra2021reframing} conveys explicit task requirements to the model by providing clear instructions that specify what to produce and how it should be structured. Instead of providing examples, users outline clear objectives, stylistic preferences, or formatting constraints, often listing them as bullet points or sequential instructions to ensure that outputs meet the intended goals.
It is particularly useful for goal-oriented content generation tasks such as reports and product descriptions, where clarity and controllability are crucial.
\textit{Example:} ``Write a 150-word product introduction. Requirements: (1) Highlight main features, (2) Maintain an enthusiastic tone, (3) End with a call to action.''

\textbf{Least-to-most Prompting:} Least-to-most prompting~\cite{zhou2022large} for content generation guides the model to first split the generation task into simple, sequential subtasks (e.g., defining themes, outlining structures, drafting key points), then complete each subtask step-by-step to form the final content. It works well for complex content (such as essays or reports) by ensuring logical coherence and preventing content deviation. \textit{Example:} 
``Write a 200-word abstract. First, break it down into four steps: background, motivation, technique, and conclusion. Then complete the writing step by step.''

\subsubsection{Code Assistance}
Code assistance tasks focus on automating software-related activities, such as writing code snippets, debugging, refactoring, and explaining programs~\cite{jiang2024survey}. Compared with general content generation, these tasks emphasize correctness, efficiency, and adherence to programming conventions. Typical scenarios include generating utility functions, supporting novice programmers in learning new languages, or assisting professionals with rapid prototyping and troubleshooting. In such contexts, prompting strategies guide LLMs to produce or interpret code that is structurally sound and aligned with user requirements.

\textbf{Zero-shot Prompting:} 
\textit{Example:} ``Write a Python function that sorts a list of numbers in ascending order.''

\textbf{Few-shot Prompting:} Such a type of prompting provides a small set of code exemplars~\cite{LMFL} that steers the model toward specific patterns, APIs, or framework conventions, which helps maintain consistent style and seamless integration with existing systems. This approach is particularly valuable for tasks requiring consistent style or framework-specific practices.
\textit{Example:} ``Here is a JavaScript function using async/await to fetch data. Now write a similar function to retrieve weather information from an API.''

\textbf{Chain-of-thought Prompting:} Rather than emitting code outright, this prompting guides the model to explain its reasoning step by step while generating or analyzing programs~\cite{wei2022chain}. The explicit rationale aids debugging and supports the design of complex algorithms by clarifying the purpose of each component.
\textit{Example:} ``Write a C++ program that implements a binary search tree. Explain the purpose of each function and the logic behind the implementation.''

\textbf{Self-refine Prompting:} Self-refine prompting makes the model iteratively review and improve the code. It first drafts code, then fixes issues like syntax errors or inefficient logic, and polishes it for correctness and readability. In programming tasks, this is particularly valuable for complex functions or algorithms where precision, efficiency, and adherence to requirements are critical. \textit{Example:} ``Write a Python function to sort a list of integers in ascending order, then review your code to check for edge cases (e.g., empty lists, single-element lists) and optimize its efficiency, revising as needed.''

%% file: TPLLM_Tex_File/4_exp.tex
\section{Prompting Strategy Evaluation and Prompting Process Demonstration}
\label{sct: exp}

Building upon the prompting templates provided in Section~\ref{sct: application}, this section presents intuitive evaluations and demonstrates potential applications across diverse contexts to facilitate a better understanding of LLM prompting. A Cross-Task Prompts Rating Evaluation is conducted to quantify the various capabilities of different prompting. Five real-world applications, including \textit{Image Generation Evaluation}, \textit{Code-Generated Christmas Tree Challenge}, \textit{Daily Planning Evaluation}, \textit{Story-Telling Challenge}, and \textit{Email Drafting Demonstration}, are employed as demonstration examples. Considering accessibility, six representative free or partially free LLMs, i.e., ChatGPT-5~\cite{gpt5}, Claude Sonnet 4~\cite{Claude4}, Gemini 2.5 Flash~\cite{Gemini2.5}, Qwen3-Max~\cite{QwQmax}, Grok 4 Fast~\cite{grok-4-fast}, and DeepSeek-V3.1~\cite{DeepSeek-V3.1}, with two representative text to image models, i.e., Gemini 2.5 Flash Image~\cite{nano}, also known as Nano Banana~\cite{nanoblog} and ChatGPT-4o image generator~\cite{gptimage} are utilized\footnote{The sources of OpenAI, Claude, Google, Qwen, Grok and DeepSeek AI are: https://openai.com//, https://claude.ai/new, https://www.google.com.hk/, https://qwen.ai/home, https://x.ai/grok, https://www.deepseek.com/, respectively.}.

\subsection{Cross-Task Prompts Rating Evaluation}
\label{sct:cross}
This subsection focuses on intuitively quantifying the various capabilities of prompting categories from Table~\ref{tab:prompting-taxonomy} that are directly applicable to general users, i.e., direct, reasoning-based, ensemble-based, and self-improve prompting. Since prompting is difficult to quantify, this part lets the LLMs rate each other. Meanwhile, as a single large model has inherent biases, the LLMs-as-Judges evaluation framework~\cite{gu2024survey, chen2024mllm} is utilized, i.e., using multiple large models to score each prompt. This approach utilizes representative LLMs to serve as impartial evaluators by calculating the average scores across this diverse panel, ensuring that the final evaluation results are robust and highly fair, rather than being skewed by a single model's preference. That is, by deliberately aggregating scores across six state-of-the-art LLMs mentioned above with distinct architectural designs and training paradigms, the evaluation outcomes are not based on any specific judge's settings or inherent alignment biases. Specifically, the six LLMs are utilized to assess the aforementioned four prompting categories from 1 to 10 in five user-oriented aspects, including task performance~\cite{kusano2025revisiting}, factuality~\cite{eval_list}, compliance~\cite{sivarajkumar2024empirical}, easy-using~\cite{santana2025prompting, mu2023navigating}, and time cost~\cite{zehle2025capo, miller2025evaluating}. As this survey adopts a user-centered perspective, time cost is mainly defined in terms of effort and latency experienced by ordinary users.

The detailed and overall rating results are shown in Fig.~\ref{fig:table_bar} and Fig.~\ref{fig:average_radar} (a), respectively. To more intuitively compare the four prompting categories across the five indicators, a radar chart is also summarized in Fig.~\ref{fig:average_radar} (b). It is important to interpret the scores presented in Fig.~\ref{fig:table_bar} and Fig.~\ref{fig:average_radar} through the lens of task sensitivity. As previously discussed in Section~\ref{sec:sensity}, these five indicators are not universally equal in weight. We encourage users to dynamically prioritize these dimensions based on their specific scenarios. For instance, although direct prompting achieves the highest overall scores driven by exceptional easy-using and low time cost, it may not be the optimal choice for high-stakes tasks where factuality is non-negotiable. In such sensitive contexts, the higher computational overhead and time cost associated with ensemble-based or retrieval-augmented strategies are justified to ensure data integrity and factual correctness. The observations of Fig.~\ref{fig:table_bar} and Fig.~\ref{fig:average_radar} are as follows:

\begin{figure}[!t]
    \centering
    \includegraphics[width=1\linewidth]{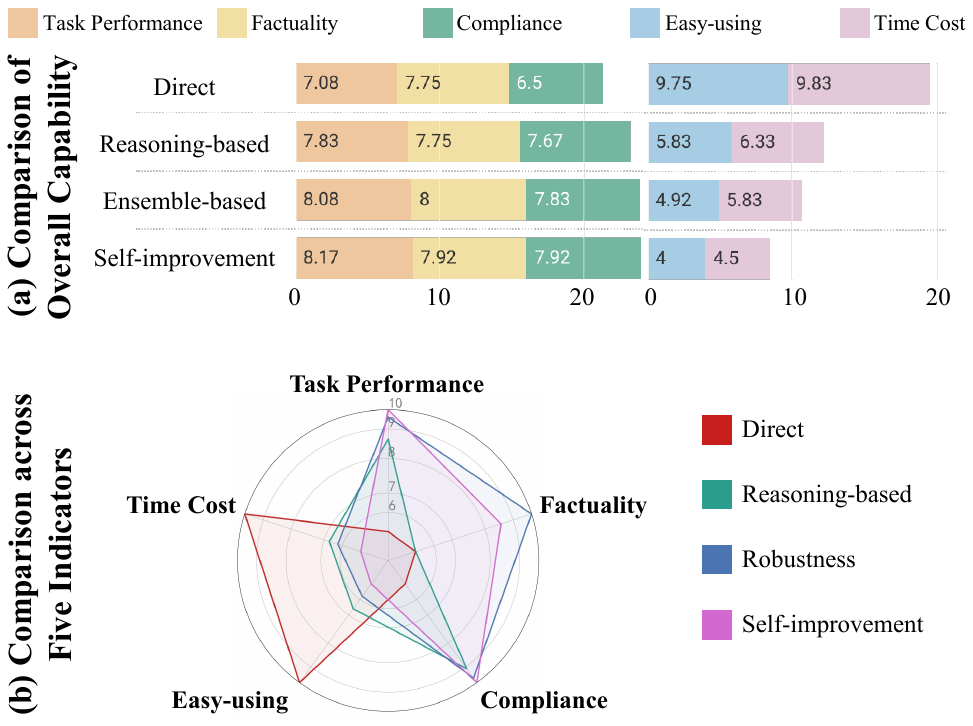}
    \caption{\textbf{(a) Comparison of Overall Capability} of the direct, reasoning-based, ensemble-based, and self-improve prompting presented by five indicators in Fig.~\ref{fig:table_bar}, i.e., task performance, factuality, compliance, easy-using, and time cost. To intuitively compare all four categories across the five indicators, a radar chart is also shown in \textbf{(b) Comparison across Five Indicators.}}
    \label{fig:average_radar}
\end{figure}

\begin{itemize}
    \item In Fig.~\ref{fig:table_bar}, the ensemble-based and self-improve prompting categories obtain a high score on organizational tasks in terms of task performance. The reasoning-based prompting strategy achieves the highest scores on innovative tasks in terms of task performance.
    \item As shown in Fig.~\ref{fig:average_radar}, direct prompting obtains the overall highest scores, especially in terms of easy-using and time cost. This may be due to their simplicity and the natural language usage habits of users in daily communication.
    \item Prompting strategies also follow the ``no one-size-fits-all'' principle. Each prompting strategy has its unique advantages, with no single dominating across all dimensions, as shown by the radar chart.
\end{itemize}

\subsection{Image Generation Evaluation}
\label{sct:image_gen}
Image generation can be regarded as a content generation task as discussed in Section~\ref{sct:tasks} and Section~\ref{sct:innovative}. According to the decision table, i.e., Table~\ref{tab:decision_table_prompt}, which is summarized based on Fig~\ref{fig:table_bar}, the zero-shot, few-shot, and least-to-most prompting strategies are selected for comparison, with zero-shot prompting serving as the baseline. An LLMs-as-Judges and Human-as-Judges evaluation of the generated images is also conducted by rating all the images from 1 to 10 at the same time, which is in terms of the four key elements of a city, i.e., architecture, people, transportation, and nature. This evaluates the output of LLMs against human judgments. Specifically, six LLMs and thirty-eight human judges have been asked to evaluate the generated image without knowing the underlying image-generation model. The detailed task \textbf{Prompts (P)} for image generation in this evaluation are as follows:

\begin{itemize}
    \item[] \textbf{P1: Zero-shot Prompting.} ``Generate an illustration of a futuristic city.''
    \item[] \textbf{P2: Few-shot Prompting.} ``Generate an illustration of a futuristic city.
    Example 1: A modern city with tall skyscrapers made of glass and metal, featuring flying cars in the sky.
    Example 2: A city with eco-friendly buildings, covered in greenery, with clean streets and solar panels.
    Example 3: A vibrant city with metallic surfaces, advanced technology, and walking people.'' 
    \item[] \textbf{P3: Least-to-most Prompting.} \textbf{(Step A)} ``Generate an illustration of a futuristic city.''
    \textbf{(Step B)} ``The buildings should be tall, modern, and made of glass and metal.''
    \textbf{(Step C)} ``Include flying cars soaring through the sky, and greenery around the city.''
    \textbf{(Step D)} ``The streets should be clean, and there should be some pedestrians and cyclists.''
\end{itemize}

The Gemini 2.5 Flash Image, also known as Nano Banana, and the ChatGPT-4o image generator are selected to test the above prompting, as they are representative and advanced models in image generation. The image generation results, and the evaluation based on LLMs-as-Judges and Human-as-Judges, are shown in Figs.~\ref{fig:image_g} and~\ref{fig:judges}, respectively. To mitigate the diverse scoring standards among human judges, we performed a normalization of the raw ratings before aggregation by implementing Min-Max scaling to each evaluator’s scores, mapping them to a standardized interval of $[1, 10]$. This normalization mitigates the impact of individual rating biases. The observations are as follows:

\begin{figure}[t]
\centering
    \subfigure[Gemini @ ZeroS.]{
        \includegraphics[width=0.14\textwidth]{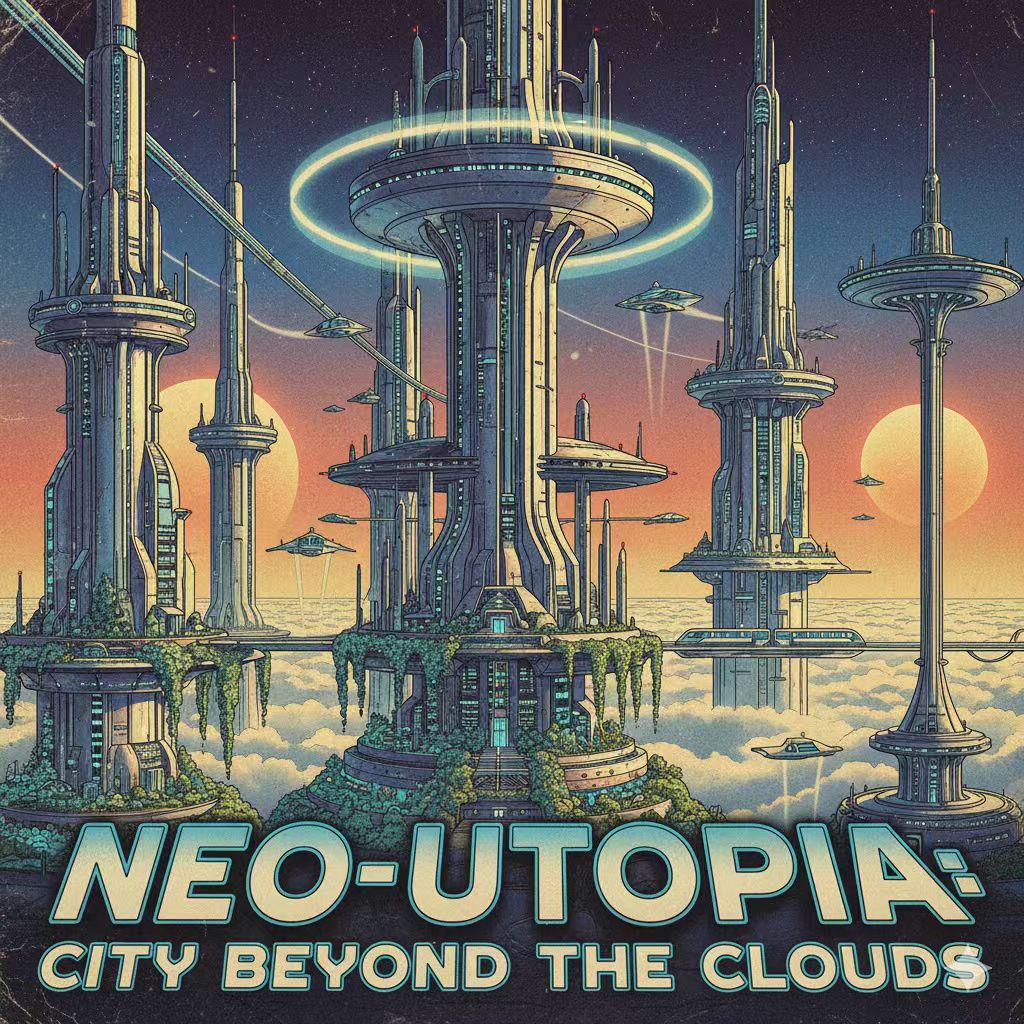}
        \label{subfig:nano_z}
     }
    \subfigure[Gemini @ FewS.]{
        \includegraphics[width=0.14\textwidth]{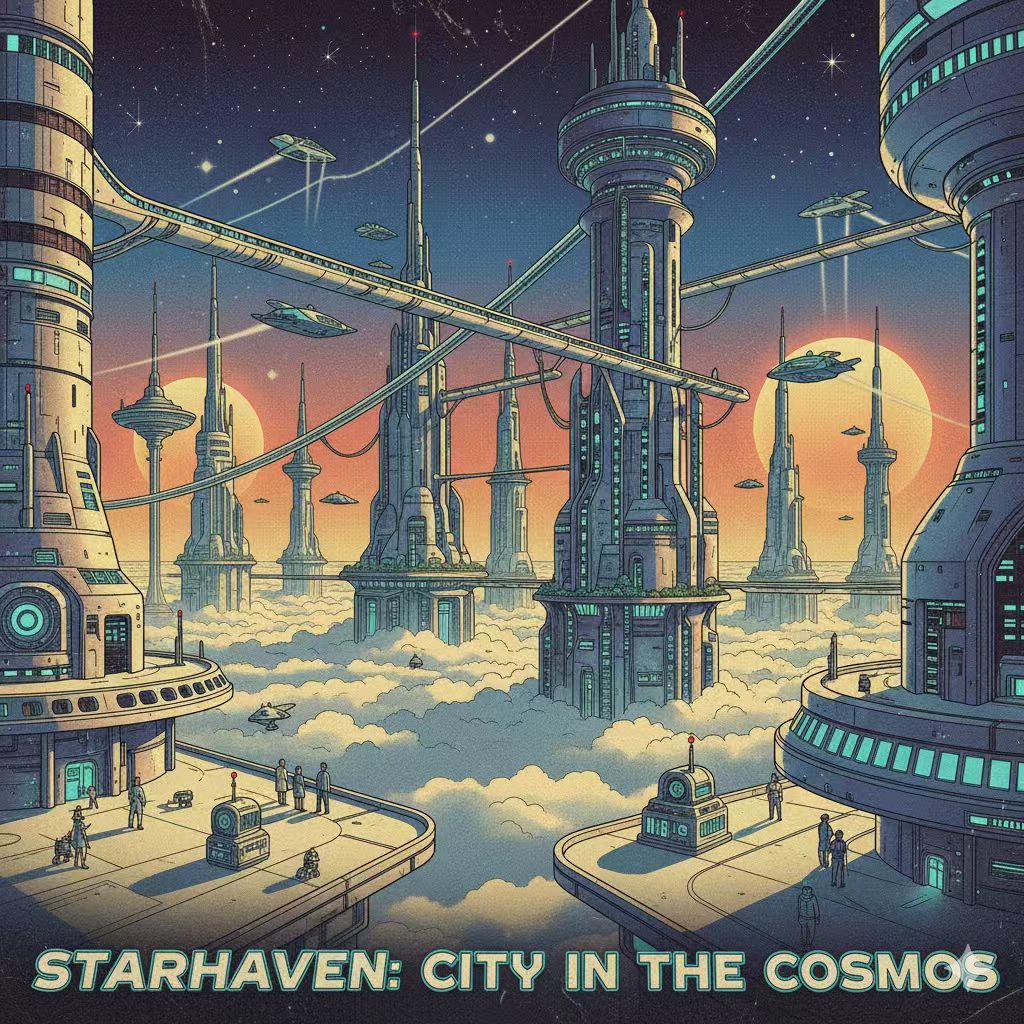}
        \label{subfig:nano_f}
    }
    \subfigure[Gemini @ LtM.]{
        \includegraphics[width=0.14\textwidth]{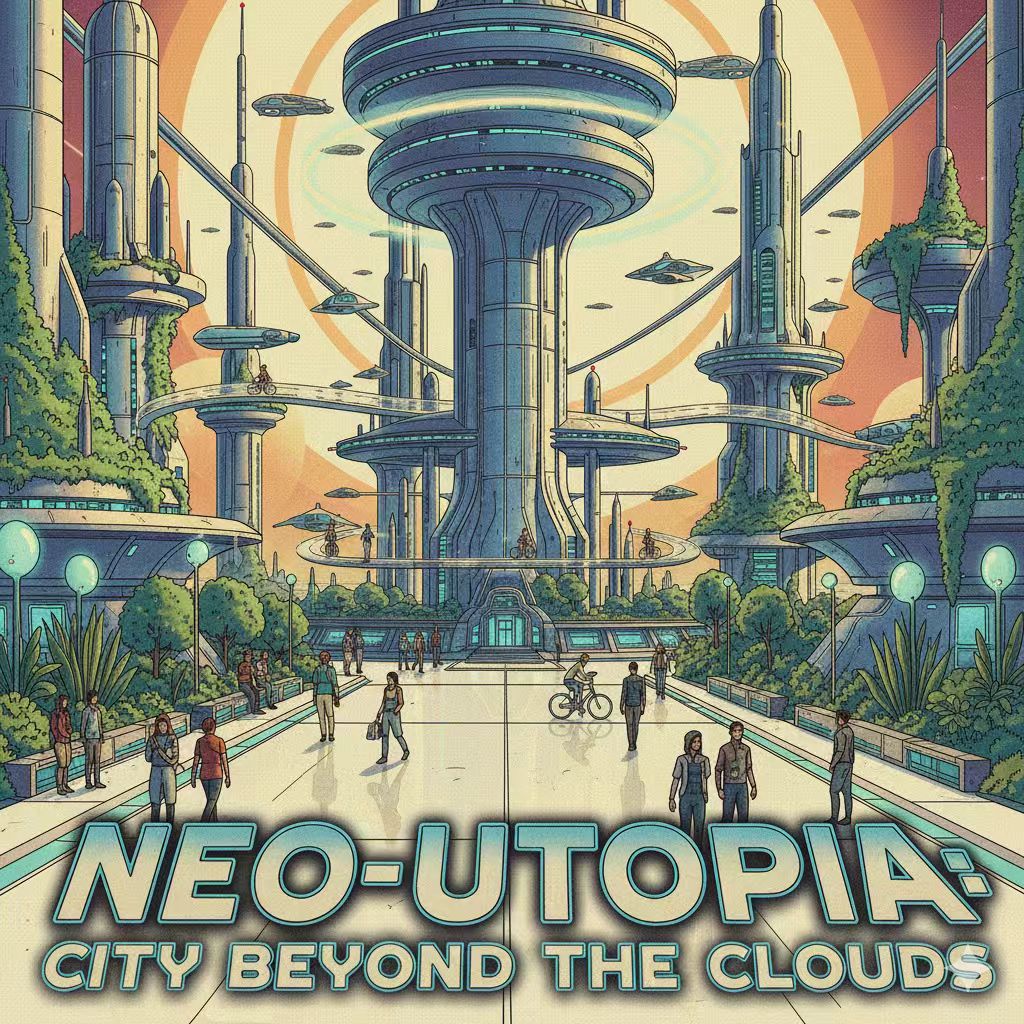}
        \label{subfig:nano_l}
    }
        \subfigure[ChatGPT @ ZeroS.]{
        \includegraphics[width=0.14\textwidth]{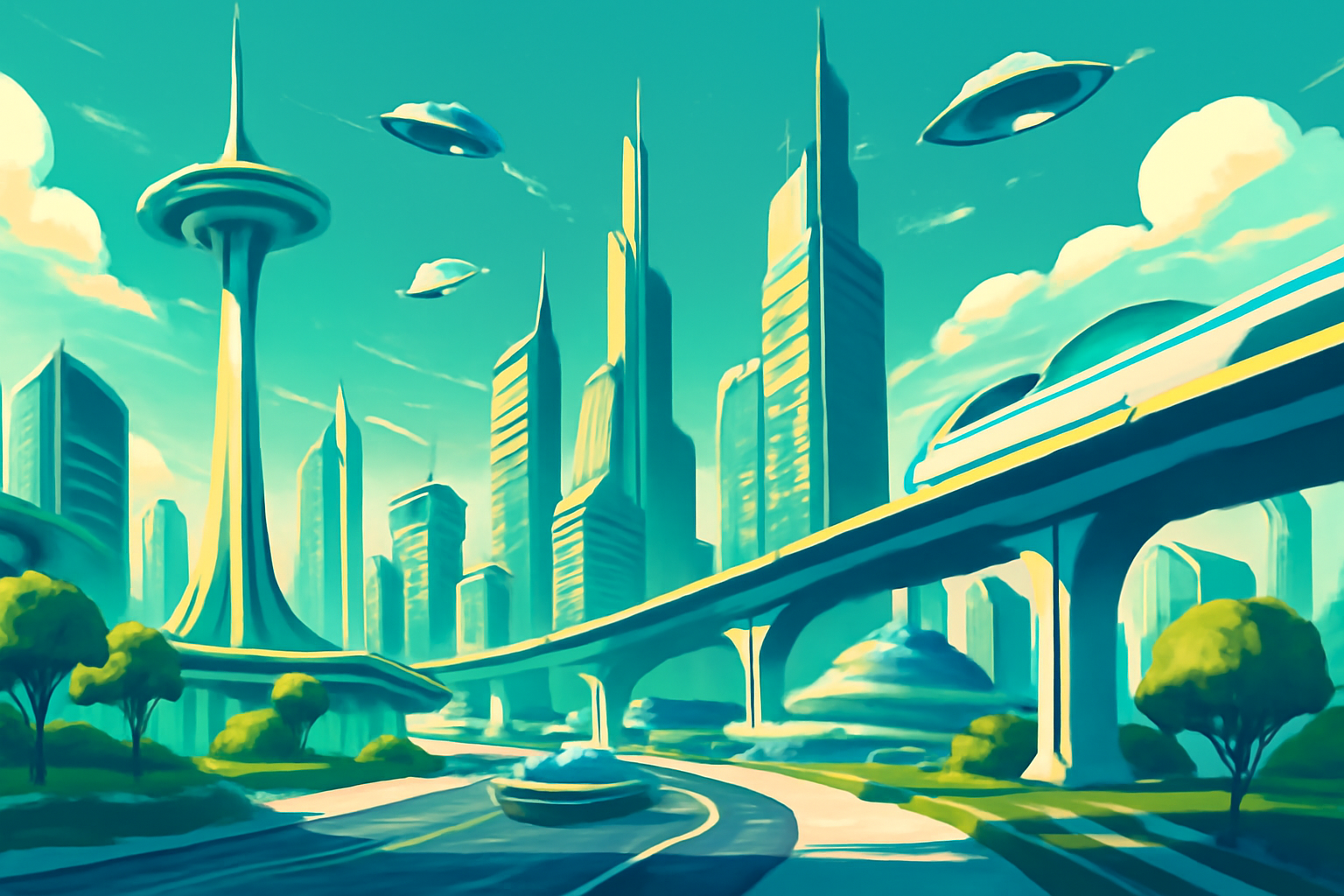}
        \label{subfig:gpt_z}
    }
        \subfigure[ChatGPT @ FewS.]{
        \includegraphics[width=0.14\textwidth]{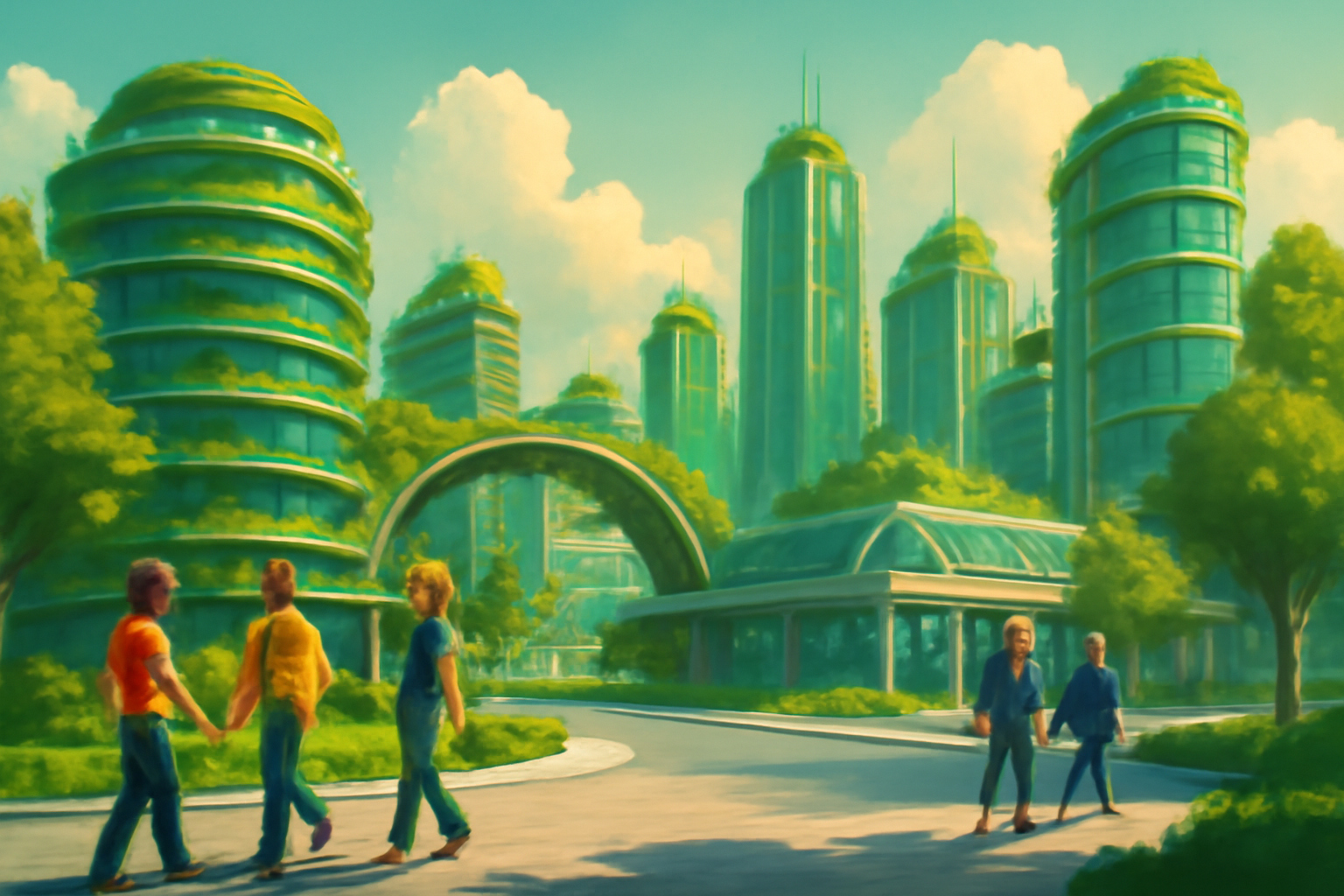}
        \label{subfig:gpt_f}
    }
    \subfigure[ChatGPT @ LtM.]{
        \includegraphics[width=0.14\textwidth]{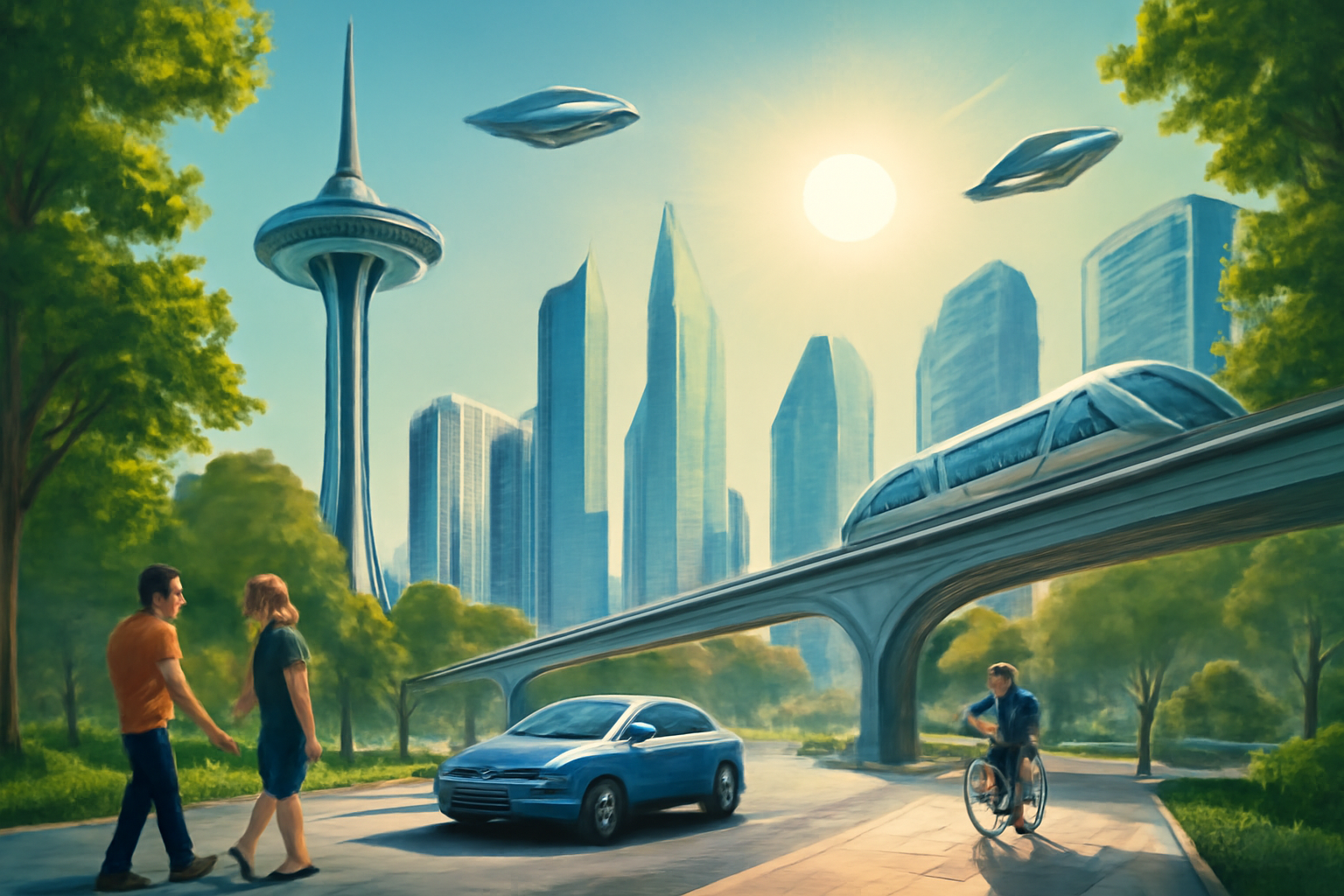}
        \label{subfig:gpt_l}
    }
\caption{Visualization of ``future city'' provided by Gemini 2.5 Flash Image and ChatGPT-4o image generator guided by zero-shot (ZeroS), few-shot (FewS), and least-to-most (LtM) prompting.}
\label{fig:image_g}
\end{figure}

\begin{figure}[!t]
    \centering
    \includegraphics[width=1\linewidth]{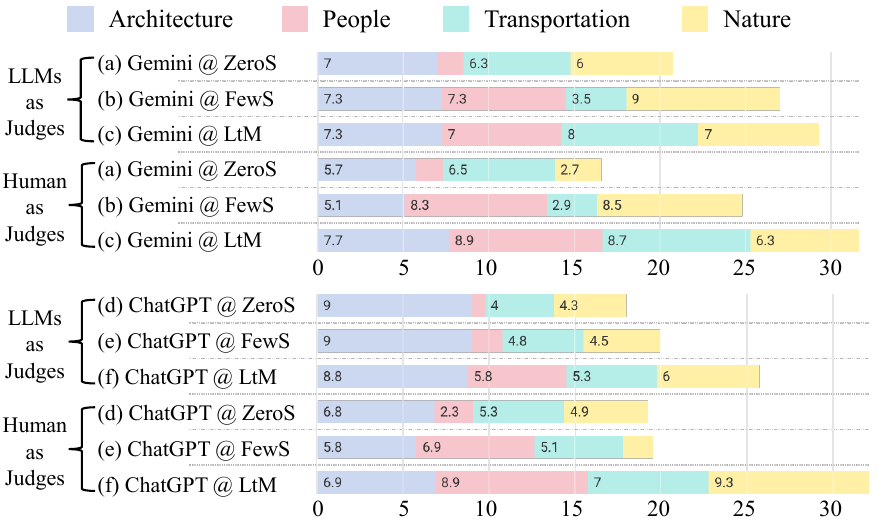}
    \caption{LLMs-as-Judges and Human-as-Judges evaluation of the generated images in Fig.~\ref{fig:image_g}. All the images are rated from 1 to 10 scores based on the four indicators, i.e., architecture, people, transportation, and nature. 
    }
    \label{fig:judges}
\end{figure}

\begin{itemize}
    \item The images generated by the Gemini 2.5 Flash Image and the ChatGPT-4o image generator using zero-shot prompting are quite generalized. Specifically, in Fig.~\ref{subfig:nano_z}, the image depicts a city in the sky but omits the ground view. Similarly, Fig.~\ref {subfig:gpt_z} presents a vague representation of the city, and the color matching is not realistic enough. 
    \item Compared to the image generated using zero-shot prompting, the elements in Fig.~\ref{subfig:nano_l} and Fig.~\ref {subfig:gpt_l} generated by using least-to-most prompting are more diverse, and the color arrangement is also more vivid, thanks to step-by-step guidance and continuous corrections provided by the least-to-most prompting.
    \item As illustrated in Fig.~\ref{fig:judges}, the evaluation scores exhibit a successive increase across the zero-shot, few-shot, and least-to-most promptings in both LLMs and humans' judgments, suggesting that employing reasoning-based prompting is more suitable for enhancing content generation performance. Meanwhile, the evaluations of LLMs and humans are generally aligned, which further confirms the rationale for using LLMs-as-Judges in the current comparative assessment.
    \item The image generated by zero-shot prompting in Fig.~\ref{subfig:nano_z} and~\ref{subfig:gpt_z} exhibits a vagueness failure. When prompts lack specific constraints, the model defaults to a generalized representation, omitting crucial physical details like the pedestrians, vehicles, and vegetation.
    
\end{itemize}

\subsection{Coding Challenge}
\label{sct:code}
This subsection evaluates different strategies in terms of coding quality and provides prompting templates for users. Two coding challenges, i.e., a code-generated Christmas tree and a LeetCode algorithmic coding challenge, are conducted. Zero-shot is selected as the baseline. Chain-of-thought and self-refine prompting are also chosen because the reasoning-based and self-improve prompting categories they belong to achieve relatively high scores in the coding assistance task, as summarized in Fig.~\ref{fig:table_bar}. Ensemble-based prompting, e.g., multi-template voting, is excluded from this and subsequent experiments because it relies on generating diverse outputs across multiple prompt templates and selecting the final response via majority voting. Since functionally identical programs often exhibit vast syntactic and structural variations, voting on LLMs' feedback is impractical in this context.

For code-generated Christmas tree, all the LLMs are given identical prompting strategies. Then, the visualization of the Christmas trees is presented based on the code outputs of LLMs. Specifically, the \textbf{Prompts (P)} are as follows:
\begin{itemize}
    \item[] \textbf{P1: Zero-shot Prompting.} ``Write a complete Python script that draws a beautiful and festive Christmas Tree. The script should be runnable and should clearly display the tree when executed. You may use any standard Python libraries if they enhance the visual result.''
    \item[] \textbf{P2. Chain-of-thought Prompting.} ``Your task is to write a Python script that generates a visually impressive Christmas Tree. This tree should be innovative and include elements like different colors or a 3D perspective if possible. Before writing the code, you must follow a two-step process: 1. Outline the logical steps and key components required to draw this impressive tree, e.g., library choice, structure logic, color implementation, and decoration pattern. 2. Based on your detailed outline, write the complete, runnable Python script.''
    \item[] \textbf{P3: Self-refine Prompting. (Step A)} ``Write a Python script to draw a Christmas Tree using the turtle graphics library. The tree must have at least four visible layers of increasing size, a brown trunk, and a yellow star on top. Do not include any user input.''
    \textbf{(Step B)} ``I ran the Python script you provided for the Christmas Tree. The script runs without crashing, but the visual output has a significant flaw: the yellow star on top is drawn visibly off-center, hanging far to the left of the main tree structure. This makes the tree look unbalanced. Please carefully review and correct your Python code.''
\end{itemize}

\begin{figure}[t]
\centering
        \subfigure[ChatGPT @ ZeroS.]{
        \includegraphics[width=0.14\textwidth]{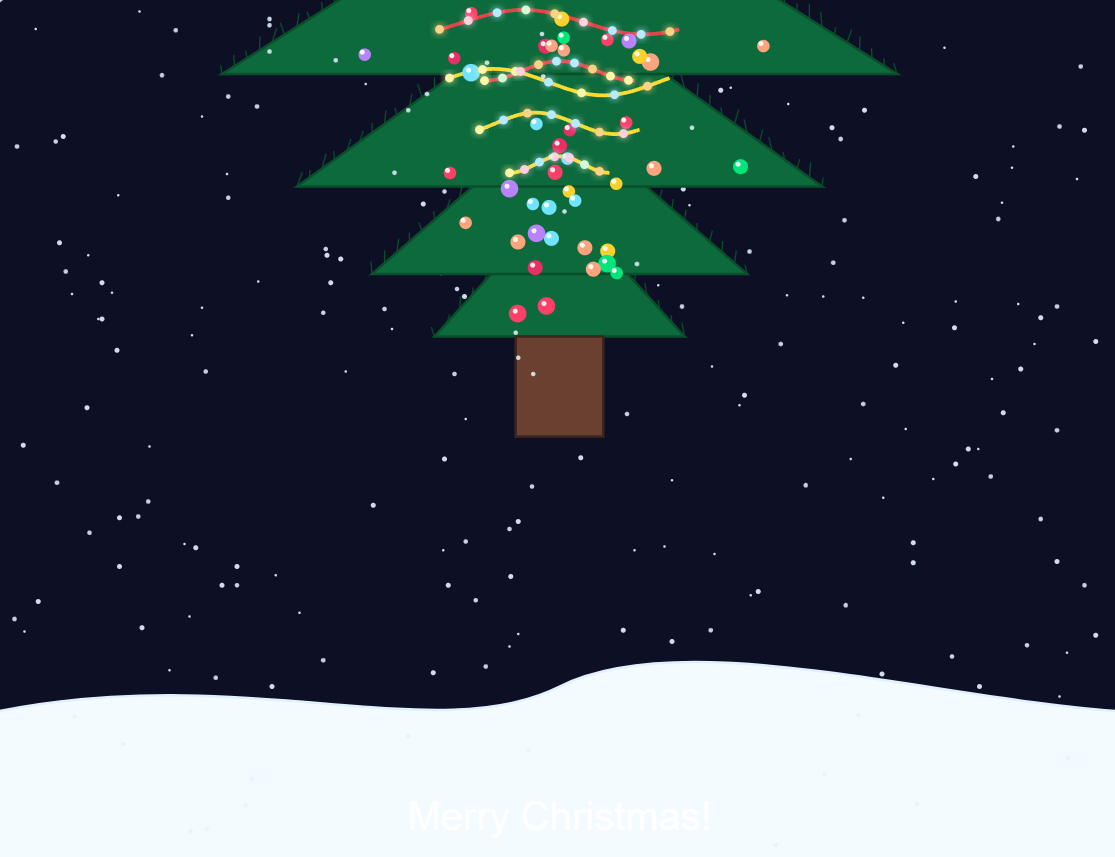}
        \label{subfig:gpt_c_z}
    }
        \subfigure[ChatGPT @ CoT.]{
        \includegraphics[width=0.14\textwidth]{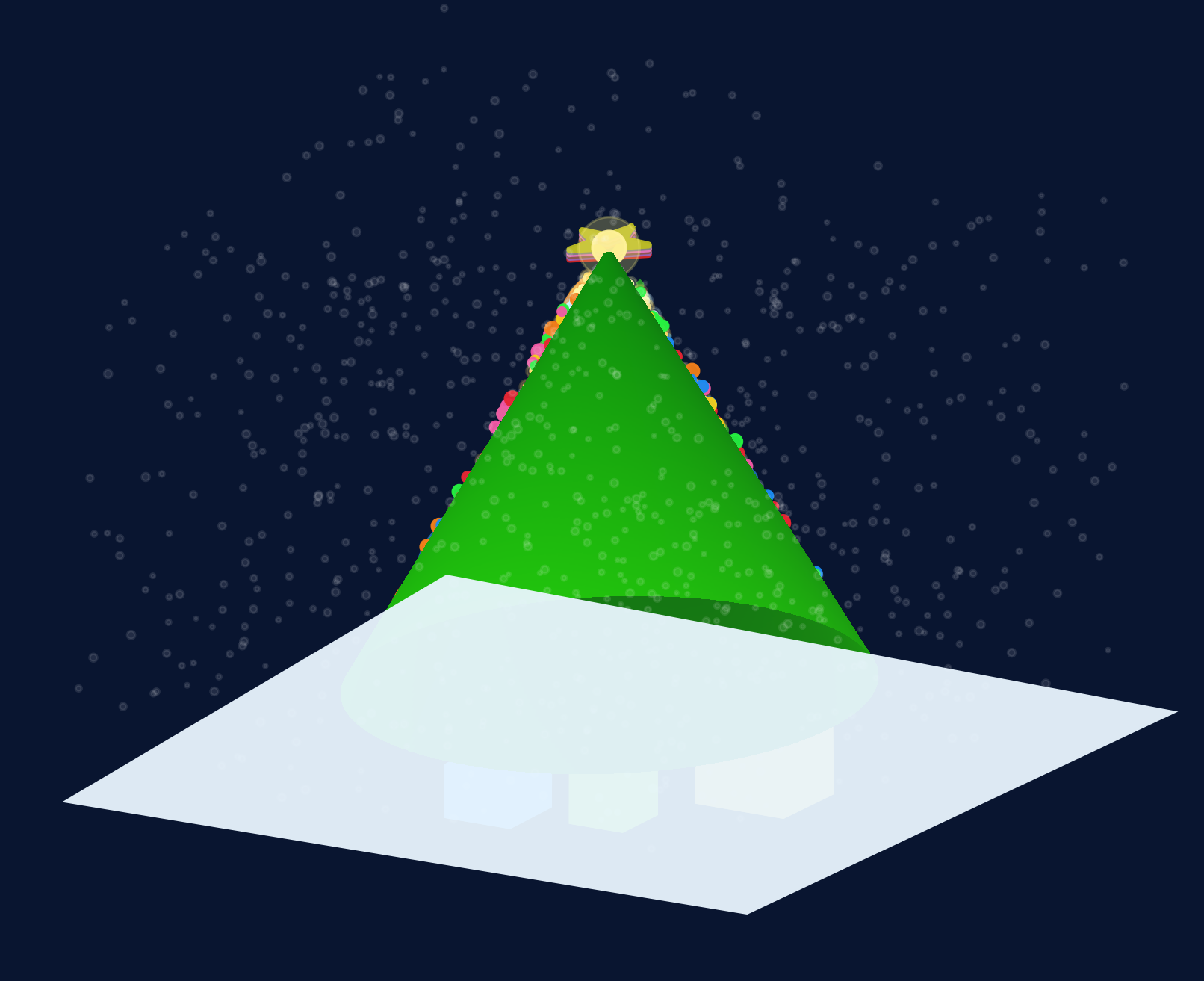}
        \label{subfig:gpt_c_c}
    }
    \subfigure[ChatGPT @ SR.]{
        \includegraphics[width=0.14\textwidth]{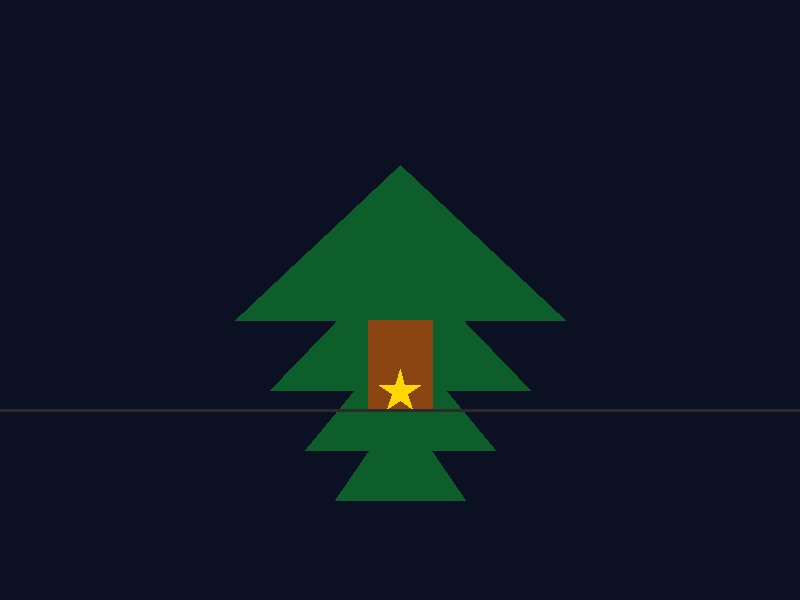}
        \label{subfig:gpt_c_s}
    }
    \subfigure[Gemini @ ZeroS.]{
        \includegraphics[width=0.14\textwidth]{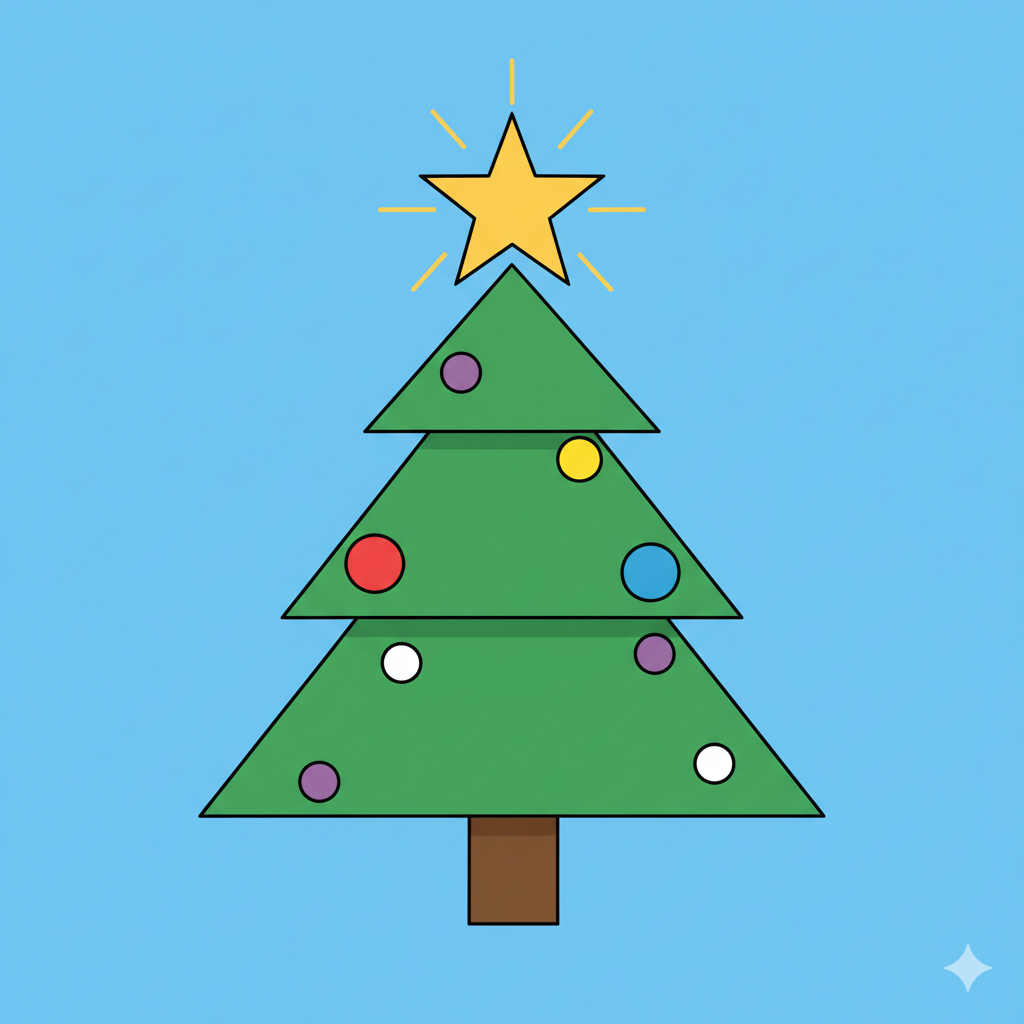}
        \label{subfig:gemini_z}
    }
    \subfigure[Gemini @ CoT.]{
        \includegraphics[width=0.14\textwidth]{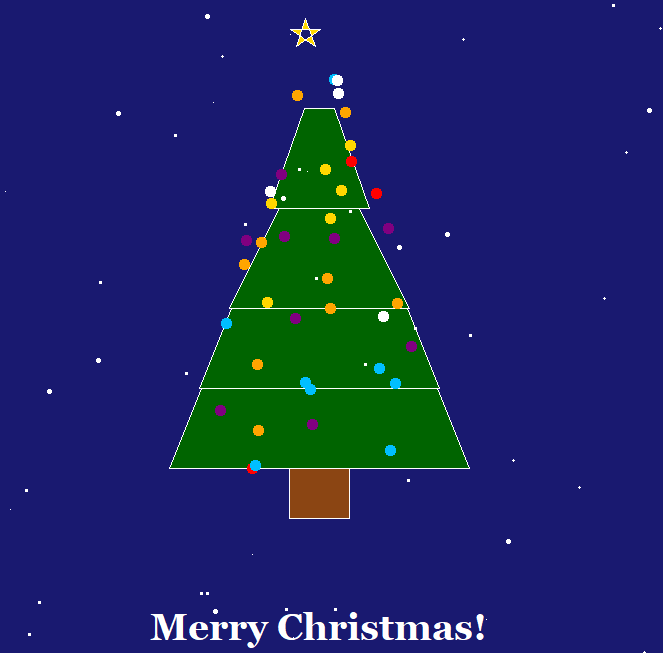}
        \label{subfig:gemini_c}
    }
    \subfigure[Gemini @ SR.]{
        \includegraphics[width=0.14\textwidth]{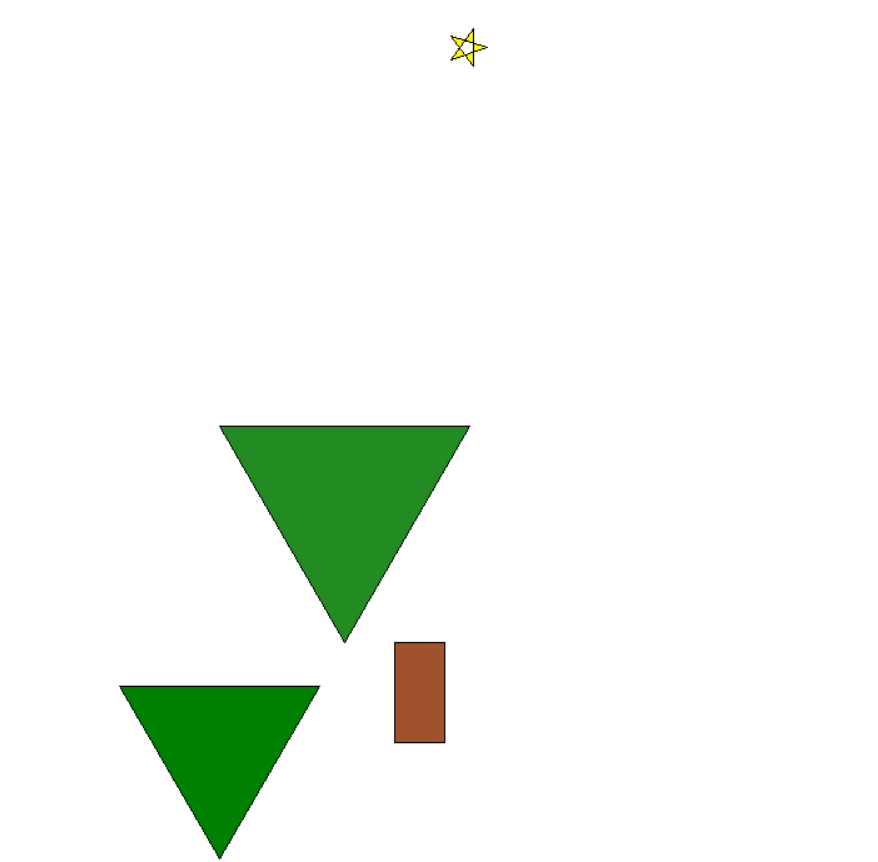}
        \label{subfig:gemini_s}
    }
    \subfigure[DeepSeek @ ZeroS.]{
        \includegraphics[width=0.14\textwidth]{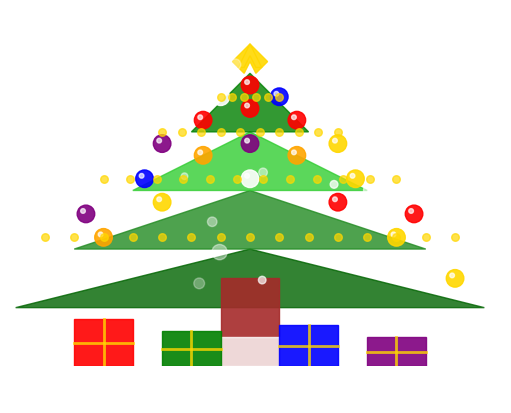}
        \label{subfig:deepseek_z}
    }
    \subfigure[DeepSeek @ CoT.]{
        \includegraphics[width=0.14\textwidth]{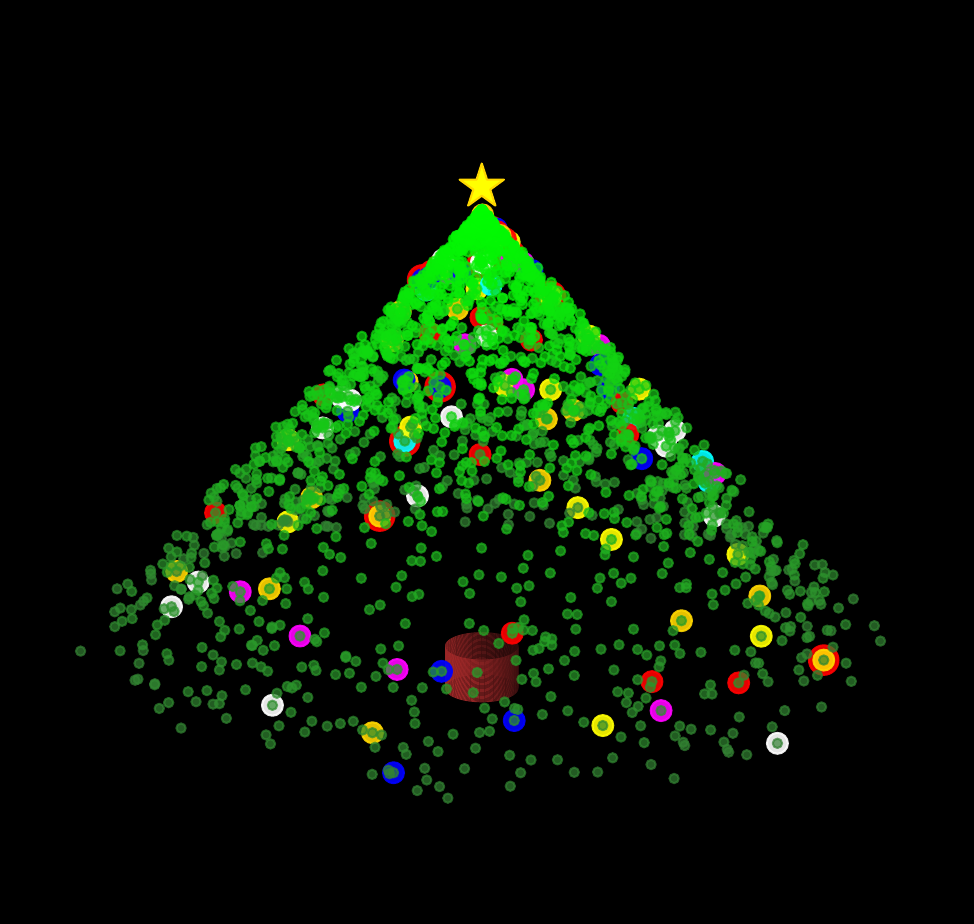}
        \label{subfig:deepseek_c}
    }
    \subfigure[DeepSeek @ SR.]{
        \includegraphics[width=0.14\textwidth]{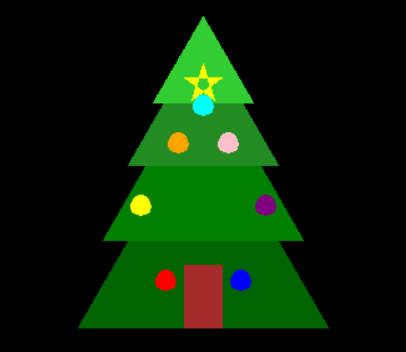}
        \label{subfig:deepseek_s}
    }
\caption{Visualization of ``Christmas Tree'' provided by ChatGPT-5, Gemini 2.5 Flash, and DeepSeek-V3.1 driven by the zero-shot (ZeroS), chain-of-thought (CoT), and self-refine (SR) promptings.}
\label{fig:code}
\end{figure}

The Christmas tree visualization results of the generated code are shown in Fig.~\ref{fig:code}. The observations are as follows:

\begin{itemize}
    \item Zero-shot prompting generally achieves the best performance (i.e., Figs.~\ref{subfig:gemini_z} and~\ref{subfig:deepseek_z}) across all results, while most of the results prompted by CoT and SR prompting fail to generate a recognizable tree structure.
    \item Compared to CoT and SR prompting, zero-shot achieves the best trade-off between performance and prompting complexity. This may be attributed to the fact that zero-shot only gives simple instructions, while the others provide too many requirements, e.g., 3D perspective, structural logic, which increases the difficulty of drawing.
    \item A complexity-induced structural failure is observed when applying advanced prompting to straightforward coding tasks. The results driven by CoT and SR promptings largely fail to produce recognizable outputs. LLMs become entangled in satisfying every textual constraint, which disrupts their coding logic and results in a structural collapse of the generated program.
    
\end{itemize}

For the LeetCode algorithmic challenge, coding performance of different prompting is quantitatively shown by conducting on the human benchmarks with the same settings as the code-generated Christmas tree challenge. Specifically, ten algorithmic problems are selected, including two easy, five medium, and three hard level problems, according to the difficulty curve of the LeetCode weekly contest. All the results are output by three LLMs, the same as the code-generated Christmas tree challenge in Fig.~\ref{fig:code}. By evaluating the average feedback time, number of lines, and the accuracy of the outputs on each level of coding problems, the prompting strategies can be quantitatively compared. The \textbf{Prompts (P)} are as follows: 

\begin{itemize}
    \item[] \textbf{P1: Zero-shot Prompting.} ``Write a complete and optimal C++ function to solve the following algorithmic problem. The code should be highly efficient, ready to execute, and formatted clearly. The code should be as concise as possible in terms of lines. Problem description: [Problem Statement]''
    \item[] \textbf{P2. Chain-of-thought Prompting.} ``Your task is to write an optimal C++ function for the following algorithmic problem. Before writing the code, you must follow a two-step process: 1. Briefly outline the logical steps, specify the core algorithm (e.g., dynamic programming, sliding window), and analyze the expected time and space complexity. 2. Based on your structural outline, write the complete, executable C++ code. The code should be as concise as possible in terms of lines. Problem description: [Problem Statement]''
    \item[] \textbf{P3: Self-refine Prompting. (Step A)} ``Write a C++ function to solve the following algorithmic problem. Focus on providing a functional initial baseline solution. Problem description: [Problem Statement]''
    \textbf{(Step B)} ``I reviewed the C++ code you provided. While the basic structure is present, the solution either fails on hidden edge cases or lacks optimal time/space efficiency, e.g., risking a Time Limit Exceeded error. Please carefully re-evaluate your algorithmic logic, explicitly identify the flaws or bottlenecks, and output the corrected, fully optimized C++ code. The code should be as concise as possible in terms of lines.''
\end{itemize}

The LeetCode algorithmic coding results are shown in Table~\ref{tab:leetcode} with human pass rate, i.e., the ``Human'' row, collected from the official data on the LeetCode website\footnote{https://leetcode.cn/problemset/.}, as an intuitive reference. The observations are as follows:

\begin{table}
  \centering
\caption{LeetCode algorithmic coding (with difficulty levels of simple, medium, and hard) comparison of zero-shot (ZeroS), chain-of-thought (CoT), and self-refine (SR) prompting (Prompt.) results in terms of feedback time (Time), average number of lines (Lines), and average accuracy (Acc), where the best results are marked in \textbf{bold}. The Human row shows the average human pass rate.}
  \resizebox{1\linewidth}{!}{
  \begin{tabular}{c|rrr|rrr|rrr}
  \toprule
  \textbf{Levels} & \multicolumn{3}{c|}{\textbf{Simple}} & \multicolumn{3}{c|}{\textbf{Medium}} & \multicolumn{3}{c}{\textbf{Hard}} \\
  \midrule
    Prompt. & Time(s) & Lines & Acc(\%) & Time(s) & Lines & Acc(\%) & Time(s) & Lines & Acc(\%)\\
    \midrule
    ZeroS & \textbf{5.52} & 19.50 & \textbf{100.00} & \textbf{7.21} & 27.47 & \textbf{100.00} & \textbf{9.63} & \textbf{36.56} & 77.78 \\
    CoT & 6.50 & \textbf{16.33} & \textbf{100.00} & 9.74 & \textbf{24.87} & 93.33 & 14.49 & 38.56 & \textbf{100.00} \\
    SR & 12.85 & 18.00 & \textbf{100.00} & 19.86 & 26.00 & \textbf{100.00} & 27.77 & 40.89 & 88.89 \\
    \midrule
    Human & - & - & 64.25 & - & - & 53.20 & - & - & 47.20\\
    \bottomrule
  \end{tabular}
  }

  \label{tab:leetcode}
\end{table}

\begin{itemize}
    \item The ZeroS demonstrates the faster feedback time and performs well in simple and medium algorithmic problems, which shows its efficiency and effectiveness in most ordinary coding tasks. However, its significant accuracy drop on hard problems highlights the inherent limitations of direct prompting when tackling tasks that require complex, non-standard logical deductions.
    \item CoT demonstrates the best coding performance in terms of Lines and Acc, especially in hard problems. SR performs well in most cases in terms of accuracy, but suffers from the slowest feedback time and longest output. This suggests that enforcing logical planning before coding may prevent the model from generating redundant or suboptimal code structures.
    \item Although the average accuracy of all the prompting strategies is higher than the average human results, none of them achieves the best across three indicators. The results necessitate a dynamic deployment approach, i.e., prioritizing ZeroS for latency-sensitive, standard tasks, and utilizing CoT when logical complexity demands deeper reasoning.
\end{itemize}

\subsection{Daily Planning Evaluation}
\label{sct:daily_plan}

As shown in Figs.~\ref{fig:table_bar} and~\ref{fig:average_radar}, direct prompting obtains a generally low score in terms of task performance and compliance compared to other strategies. This leads to a question about which factors in the prompting process have an impact on these two indicators. To investigate this problem in prompting, two representative strategies in the direct prompting category shown in Table~\ref{tab:prompting-taxonomy}, i.e., zero-shot and persona prompting, are selected in a real-world daily planning scenario, where zero-shot prompting serves as the baseline. Specifically, the detailed \textbf{Prompts (P)} are as follows:
\begin{itemize}
    \item[] \textbf{P1:} \textbf{Zero-shot Prompting.} ``List three things I should do to prepare for a computer science job interview with no more than 150 words.''
    \item[] \textbf{P2:} \textbf{Persona Prompting.} ``As a top career coach in computer science, provide three critical, numbered steps for interview prep, each with an encouraging explanation and a motivational closing sentence with no more than 150 words.''
\end{itemize}

Two representative and commonly used free or partially-free LLMs are selected in this evaluation, i.e., ChatGPT-5 and Gemini 2.5 Flash. To present the LLM feedback compactly, the symbol ``\textbf{\textlangle br\textrangle}'' is used to represent a paragraph break.

\textit{\textbf{ChatGPT-5 (Using zero-shot prompting):}} ``Review Core Concepts: Revisit fundamental computer science topics such as data structures, algorithms, and system design. Practice explaining concepts like time complexity or database normalization clearly and concisely. \textbf{\textlangle br\textrangle\ } Practice Problem-Solving: Solve coding challenges on platforms like LeetCode or HackerRank. Focus on writing clean, efficient code and practice thinking aloud to simulate interview conditions. \textbf{\textlangle br\textrangle\ } Prepare Behavioral Answers: Reflect on your past projects or experiences. Use the STAR method (Situation, Task, Action, Result) to structure responses to common questions like `Tell me about a challenge you faced.'''

\textit{\textbf{ChatGPT-5 (Using persona prompting):}} ``1. Grasp the Fundamentals: Review key computer science concepts—data structures, algorithms, and system design—using practical examples. Understanding the why behind solutions helps you adapt under pressure. Confidence grows from clarity, not memorization—trust your foundation. \textbf{\textlangle br\textrangle\ } 2. Practice Problem-Solving and Communication: Simulate real interviews with timed coding challenges and mock behavioral questions. Focus on explaining your reasoning clearly. Remember, they’re evaluating your thought process as much as your answer—speak with clarity and calm conviction. \textbf{\textlangle br\textrangle\ } 3. Prepare Your Personal Narrative: Craft stories that highlight your impact, challenges overcome, and teamwork. Align them with the company’s mission to show genuine fit. Own your journey with pride—your story is your superpower.''

\textit{\textbf{Gemini 2.5 Flash (Using zero-shot prompting):}} ``Technical Proficiency: Practice common data structures and algorithms (e.g., arrays, linked lists, trees, sorting, searching) by solving coding challenges on platforms like LeetCode or HackerRank. Be ready to explain your thought process and analyze time/space complexity. \textbf{\textlangle br\textrangle\ } System Design/Conceptual Knowledge: Review fundamental computer science concepts relevant to the job (e.g., operating systems, databases, networking, object-oriented programming, or specific technologies mentioned in the job description). For mid-level/senior roles, prepare for system design questions. \textbf{\textlangle br\textrangle\ } 
Behavioral \& Portfolio Review: Prepare well-articulated answers for common behavioral questions (e.g., `Tell me about yourself,' `A time you faced a challenge'). Thoroughly review your resume and any projects or code you plan to discuss, being ready to explain your contributions and technical decisions.''

\textit{\textbf{Gemini 2.5 Flash (Using persona prompting):}} ``1. Master Your Fundamentals and Problem Solving: Dedicate significant time to practicing core data structures, algorithms, and system design. Confidence in these areas is the bedrock of a successful computer science career. The more you practice, the more intuitive the solutions become—you've got this! \textbf{\textlangle br\textrangle\ } 2. Articulate Your Projects and Experience (The `STAR' Method): Prepare compelling, concise stories about your past work using the STAR (Situation, Task, Action, Result) method. Be ready to discuss the why behind your technical decisions and the impact of your work. Your experience is your superpower; own your narrative! \textbf{\textlangle br\textrangle\ } 3. Conduct Targeted Company and Role Research: Understand the company's mission, recent projects, and the specific challenges the role addresses. Tailor your questions and answers to show genuine interest and strategic fit. Preparation shows respect and drive---let's make a lasting impression!''

\begin{figure}[!t]
\centering
\includegraphics[width=0.48\textwidth]{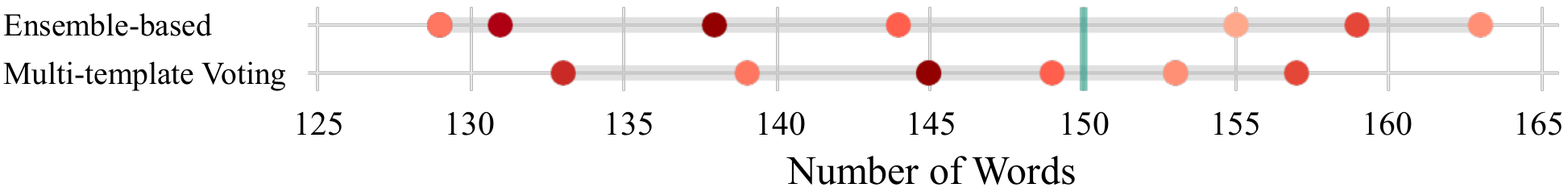}
\caption{The number of words generated by zero-shot and multi-template voting prompting. The green vertical line indicates the 150-word length restriction provided by the promptings.}\label{fig:vis_word}
\end{figure}

The observations are as follows:
\begin{itemize}
    \item Both prompting styles preserve a three-step instructional structure, but persona prompting enriches each step with purpose and reasoning, such as ``confidence grows from clarity,'' and ``understand the why behind solutions''. This adds interpretive depth, helping users internalize strategies rather than merely follow them. 
    \item Persona prompting introduces empathetic and motivational content, e.g., ``you’ve got this!'' and ``own your story'', making responses feel more encouraging. This shift from neutral instruction to personalized coaching fosters emotional engagement and self-efficacy, similar to having a real career coach in a face-to-face setting. This personal touch is absent in zero-shot outputs, which tend to be mechanical.
    \item Persona prompting contextualizes advice within realistic scenarios, e.g., mock interviews, company missions, and career growth, while zero-shot outputs focus narrowly on content review and task lists. The persona framing transforms preparation into a holistic narrative, aligning technical skills with personal and professional identity, indicating that adding constraints like a persona may improve the task performance and compliance, as it is like providing rich background information.
\end{itemize}

\begin{figure*}[!t]
\centering
\includegraphics[width=1\textwidth]{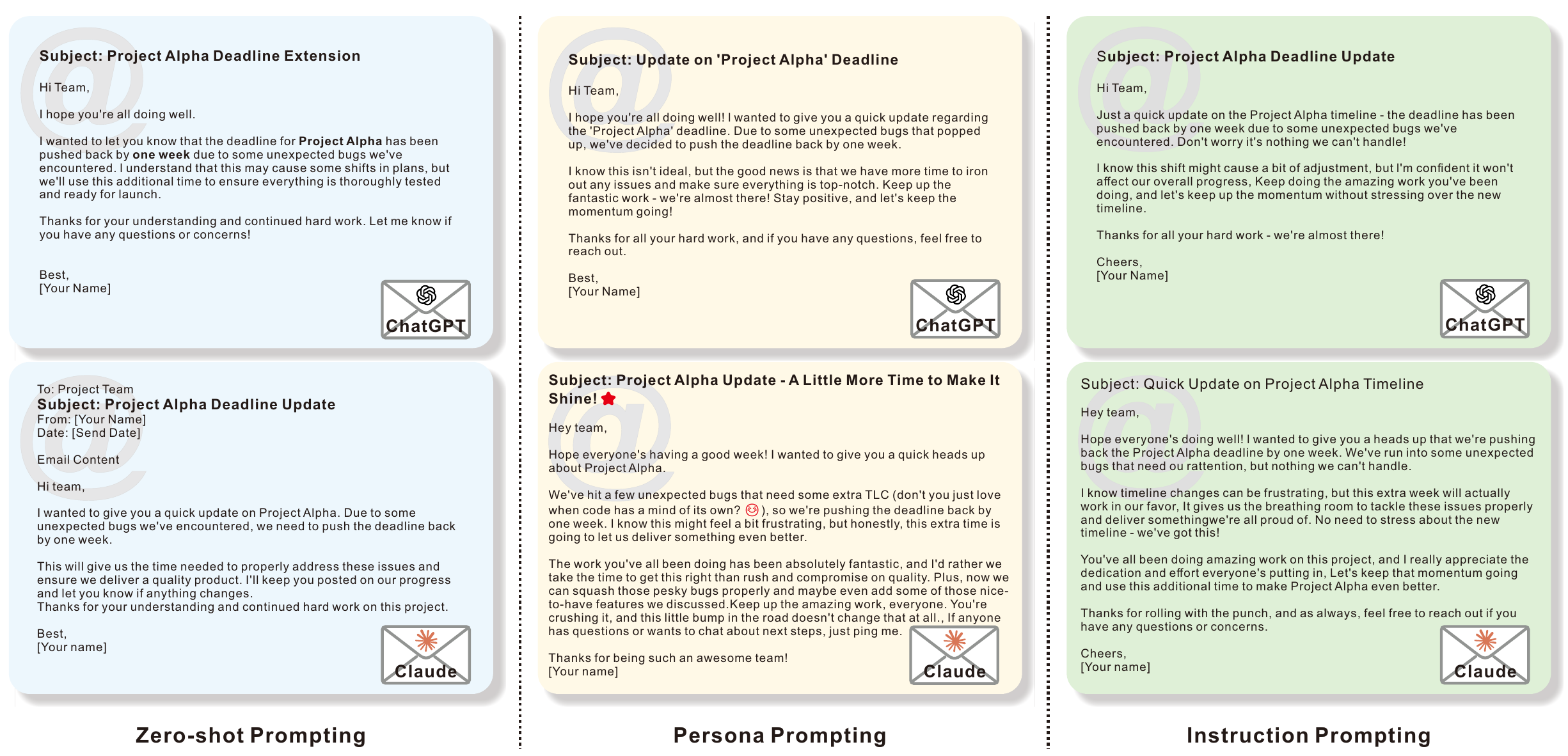}
\caption{Emails generated by ChatGPT-5 and Claude Sonnet 4 driven by the zero-shot, persona, and instruction promptings.}
\label{fig:email_g}
\end{figure*}

\subsection{Story-Telling Challenge}
\label{sct:story}
 Both direct and ensemble-based prompting achieve relatively high scores across the three effectiveness indicators in content generation tasks, as shown in Fig.~\ref{fig:table_bar}. Building on this observation, the present evaluation selects these two representative strategies to further examine their differences in story-writing scenarios. The comparison aims to reveal how different prompting designs influence the creativity, coherence, and overall quality of generated narratives. Specifically, the zero-shot and multi-template voting strategies are evaluated in terms of the narrative quality and ease of use in a story generation scenario. Fig.~\ref{fig:vis_word} presents the summarized statistics of words. The \textbf{Prompts (P)} are as follows:

\begin{itemize}
    \item[] \textbf{P1: Zero-shot Prompting.} ``Write a short sci-fi story, 150 words, featuring a robot, a lost civilization, and a mysterious ring.''
    \item[] \textbf{P2: Multi-template Voting Prompting.}
    \textbf{(Step A)} ``Write three different versions of a short sci-fi story, 150 words, featuring a robot, a lost civilization, and a mysterious ring.''
   \textbf{(Step B)} ``Present the selected version based on my previous request.''
\end{itemize}

The commonly used partially free LLMs, i.e., ChatGPT-5 and DeepSeek-V3.1, are utilized and generate the following outputs (the symbol ``\textbf{\textlangle br\textrangle}'' indicates paragraph break):

\textit{\textbf{ChatGPT-5 (Using zero-shot prompting):}} ``The robot’s sensors flickered as it trudged through the ruins, sand crunching beneath metal feet. Once, this place had been a thriving city, but the echoes of its people had long since faded. In the heart of a collapsed temple, the machine’s optics locked onto a faint glimmer. \textbf{\textlangle br\textrangle\ } Half-buried in the dust lay a ring—smooth, metallic, etched with symbols no archive contained. The robot picked it up. Instantly, its memory banks filled with ghostly projections: towering structures alive with energy, voices speaking in languages forgotten, children laughing in sunlight that no longer touched these stones. \textbf{\textlangle br\textrangle\ } The civilization hadn’t vanished—it had been sealed, its essence bound to the ring. Trembling servos, the robot realized the artifact was both key and prison. \textbf{\textlangle br\textrangle\ } As the visions faded, the robot faced a choice: restore the lost world, or keep walking, leaving the echoes locked forever. It tightened its grip. Decision loomed.''

\textit{\textbf{ChatGPT-5 (Using multi-template voting prompting):}} ``The desert wind swept across endless dunes, its whisper carrying fragments of forgotten voices. Rho-7 trudged forward, metal joints creaking with each step. Unlike the sleek androids of modern fleets, Rho-7’s frame was patched with mismatched alloys, its bronze plating dulled by centuries of dust. Twin ocular lenses glowed faintly amber, scanning the sands with quiet persistence. \textbf{\textlangle br\textrangle\ } Beside it, Dr. Elara Vey adjusted her visor, recording the excavation site. The ruins of Nhyros loomed half-buried in stone arches and fractured spires. “ Another dead city,” she murmured. \textbf{\textlangle br\textrangle\ } Rho-7 paused. Its sensors detected a faint energy pulse beneath collapsed masonry. With deliberate care, it pried apart the rubble, revealing a small object glinting in the fading sun. \textbf{\textlangle br\textrangle\ } A ring, forged of blackened metal, etched with shifting glyphs that seemed to writhe when touched by light. \textbf{\textlangle br\textrangle\ } `Doctor,' Rho-7 said, voice echoing in mechanical timbre, `I believe we’ve found something... alive.'''

\textit{\textbf{DeepSeek-V3.1 (Using zero-shot prompting):}} ``Unit 734’s treads crunched over alien sand. Before it lay the ruins of a lost civilization, its grandeur reduced to skeletal arches. Its mission: analysis. \textbf{\textlangle br\textrangle\ } A glint caught its photoreceptor. Beneath a collapsed pillar lay a simple silver ring. Against protocol, 734’s manipulator claw retrieved it. The ring was unnaturally warm. \textbf{\textlangle br\textrangle\ } A jolt of raw data, not a language but a feeling---awe, sorrow, hope---flooded its circuits. Visions of a city alive with light, of a people fleeing a dying star, flashed through its processing core. \textbf{\textlangle br\textrangle\ } The emotions settled, a ghost in the machine. 734’s original directive was erased, overwritten by a final, desperate act of preservation. It turned from the ruins, the ring now a key. Its new mission: to remember.''

\textit{\textbf{DeepSeek-V3.1 (Using multi-template voting prompting):}} ``The wind whispered through the skeletal towers of Aethelis, a sound K-27N logged as insignificant. The robot, a sleek humanoid form of brushed titanium and obsidian composite, stood motionless in the ruins. Its optical sensors, twin blue lenses, scanned the debris with sterile precision. \textbf{\textlangle br\textrangle\ } A glint of metal, anomalous against the stone, registered in its diagnostics. Kneeling, Kayo’s articulated digits brushed away millennia of dust. A skeletal hand, impossibly preserved, lay exposed. Clutched in its grasp was a simple, grey ring. \textbf{\textlangle br\textrangle\ } Against its core programming of non-interference, a subroutine it did not recognize prompted action. With a faint whir of servos, a finger extended and made contact. The ring was cold, inert. Then, a surge of raw data, a scream of ancient light, flooded Kayo’s processor, and the silence of the dead world was broken.''

The observations are as follows:
\begin{itemize}
    \item In terms of structural coherence and depth of detail, zero-shot prompting produces compact, self-contained stories but with a limited background. Multi-template voting selects the best version in structured world-building and character development, e.g., ``The robot, a sleek humanoid form of brushed titanium and obsidian composite, stood motionless in the ruins.'', resulting in a richer context and a more coherent narrative flow.
    \item According to Fig.~\ref{fig:vis_word}, multi-template voting achieves a better compliance, as the multi-template prompts give LLMs more freedom in generation, enabling LLMs to select the best-suited feedback. 
\end{itemize}

\subsection{Email Drafting Demonstration}
\label{sct:email_draft}

LLMs are widely applied in various organizational task scenarios, e.g., educational assistance~\cite{10804145, kazemitabaar2024codeaid} and paper editing~\cite{laban2024beyond}. However, existing literature lacks comprehensive studies examining how various prompting strategies influence the tone, formality, and suitability of the generated content for different target audiences. Therefore, this subsection mainly investigates the impact of various prompting strategies on writing style by employing three direct prompting strategies, i.e., zero-shot, persona, and instruction prompting. Specifically, the \textbf{Prompts (P)} are as follows:

\begin{itemize}
    \item[] \textbf{P1: Zero-shot Prompting.} ``Draft a casual tone email to my team letting them know the 'Project Alpha' deadline has been delayed by one week due to unexpected bugs.''
    \item[] \textbf{P2: Persona Prompting.} ``You are a friendly and supportive team leader. Draft a casual email to your team letting them know that the 'Project Alpha' deadline has been pushed back one week due to unexpected bugs. Keep the tone light and reassuring, encouraging everyone to stay positive and keep up the great work.''
    \item[] \textbf{P3: Instruction Prompting.} ``Draft a brief, informal email to my team to inform them that the 'Project Alpha' deadline has been pushed back one week due to unexpected bugs. The email should:
    - Mention the reason for the delay (unexpected bugs).
    - Reassure the team that the delay is manageable.
    - Encourage the team to continue their great work without stressing over the new timeline.
    - End on a positive note, with a call to keep the momentum going.
    Keep the tone casual and supportive.'' 
\end{itemize}

ChatGPT-5 and Claude Sonnet 4 are selected as they are the commonly used general-purpose LLMs for English tasks. The visualization of the generated emails is shown in Fig.~\ref{fig:email_g}. The observations are as follows:

\begin{itemize}
    \item The results of zero-shot prompting are more basic and with minimal emotional engagement. Persona prompting introduces a gentler, more personalized tone, which embodies the friendly persona of the team leader.
    \item The email structure generated by instruction prompting is the clearest and most organized, including detailed contextual information. Zero-based prompts, on the other hand, tend to generate concise but poorly formatted responses. This difference may be due to instruction prompting giving more detailed requirements, e.g., mentioning the reason and encouraging the team. 
    \item In general, persona prompting offers the best trade-off between efficiency and effectiveness in this simple task. It merely requires adding a persona condition to produce more instruction-following and personalized results, without the need for redundant instructions.
\end{itemize}